\documentclass[preprint, onecolumn, secnumarabic, nobalancelastpage, amsmath, superscriptaddress, notitlepage, tightenlines]{revtex4-2}

\usepackage[T1]{fontenc}
\usepackage[latin9]{inputenc}
\usepackage{amstext}
\usepackage{amssymb}
\usepackage{graphicx}
\usepackage{wasysym} 

\usepackage{sidecap}

\newcommand{\Azo}{\textit{Azoarcus}}
\newcommand{\Tet}{\textit{Tetrahymena}}
\newcommand{\Mg}{Mg\textsuperscript{2+}}

\newcommand{\K}{K\textsuperscript{+}}

\newcommand{\Rg}{\ensuremath{R_{\textrm{g}}}}
\newcommand{\meanRg}{\ensuremath{\left.\langle R_{\textrm{g}} \rangle \right.}}
\newcommand{\Ic}{$\textrm{I}_{\textrm{c}}$}

\usepackage{xcolor}
\usepackage{soul}

\begin{document}

\title{Watching ion-driven kinetics of ribozyme folding and misfolding caused by energetic and topological frustration one molecule at a time}

\author{Naoto Hori}
\email{hori.naoto@gmail.com}
\affiliation{Department of Chemistry, University of Texas, Austin, TX 78712, USA}
\affiliation{School of Pharmacy, University of Nottingham, Nottingham, UK}
\author{D. Thirumalai}
\email{dave.thirumalai@gmail.com}
\affiliation{Department of Chemistry, University of Texas, Austin, TX 78712, USA}
\affiliation{Department of Physics, University of Texas, Austin, TX 78712, USA}

\date{Aug 21, 2023}

\begin{abstract}
Folding of ribozymes into well-defined tertiary structures usually requires divalent cations. How \Mg{} ions direct the folding kinetics has been a long-standing unsolved problem because experiments cannot detect the positions and dynamics of ions. To address this problem, we used molecular simulations to dissect the folding kinetics of the \Azo{} ribozyme by monitoring the path each molecule takes to reach the folded state. We quantitatively establish that \Mg{} binding to specific sites, coupled with counter-ion release of monovalent cations, stimulate the formation of secondary and tertiary structures, leading to diverse pathways that include direct rapid folding and trapping in misfolded structures. In some molecules, key tertiary structural elements form when \Mg{} ions bind to specific RNA sites at the earliest stages of the folding, leading to specific collapse and rapid folding. In others, the formation of non-native base pairs, whose rearrangement is needed to reach the folded state, is the rate-limiting step. Escape from energetic traps, driven by thermal fluctuations, occurs readily. In contrast, the transition to the native state from long-lived topologically trapped native-like metastable states is extremely slow. Specific collapse and formation of energetically or topologically frustrated states occur early in the assembly process.
\end{abstract}

\maketitle

\includegraphics[width=0.6\textwidth]{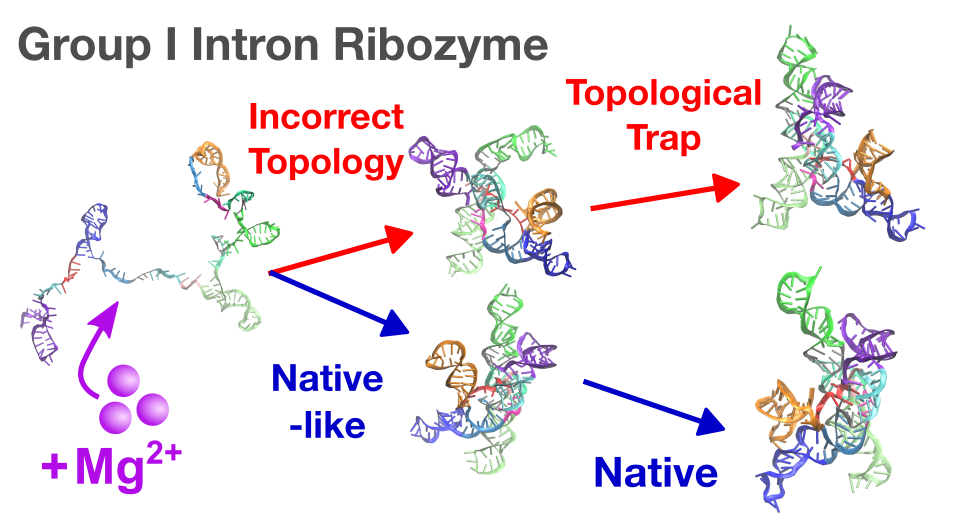}

\textbf{Graphical abstract}

\section*{Introduction}
Many functional RNA molecules fold to specific tertiary structures, in which divalent cations play crucial roles \citep{HeilmanMillerI01JMB, Grilley06PNAS, Koculi07JACS, Bowman12COSB, Lipfert14ARBiochem, Sun17ARBiophys,Thirumalai05Biochem}. Their folding pathways often consist of multiple steps covering a spectrum of time scales, and traversal through multiple pathways \citep{Pan97JMB, Roca18PNAS}. The structural ensemble of folding intermediates is, therefore, highly heterogeneous \citep{Xie04P}. In spite of advances in experimental methods, there are still limitations to the spatial and temporal resolution in dissecting the folding of large RNA molecules, especially the mechanisms by which divalent cations modulate the folding landscape.

Group I intron is a self-splicing ribozyme, that has been widely used in studies of RNA folding, typically either from purple bacteria \Azo{} or ciliates \Tet{} \cite{VicensCech06}. In the presence of divalent cations (\Mg{}), the RNA molecule folds to a specific tertiary structure (Fig.~\ref{fig:Fig1}a-b) \citep{AzoXray04N}. It is known that the folding takes place in a hierarchical manner \citep{Woodson10ARBiophys,Roh10JACS}. \Mg{} is indispensable not only for its catalytic activity but also for the formation of the native conformation \citep{Rook99PNAS, Thirumalai00RNA}. The double-strand helices in the core region can fold at \Mg{} $\sim$0.2 mM, whereas the complete tertiary structure requires at least $\sim$2 mM \Mg{}. Woodson and coworkers revealed, by time-resolved Small Angle X-ray Scattering (tSAXS) experiments, up to 80\% fraction of unfolded \Azo{} ribozyme reach compact structures in less than 1 ms upon the addition of 5 mM \Mg{} \citep{Roh10JACS}. The overall folding time has been estimated to be 5 to 50 ms for the major fraction of RNA in ensemble experiments, which is faster in \Azo{} than \Tet{} ribozyme \citep{Chauhan08JACS}, because \Azo{} ribozyme has smaller and simpler peripheral domains. On the other hand, it has also been known that a certain fraction of the molecule is trapped in an intermediate state, presumably because of misfolding. This persistent intermediate state needs times on the order of minutes to hours to fold to the native conformation \citep{Sinan11JBC}.

Here, we simulate the multistep folding kinetics of \Azo{} group I intron by extensive Brownian dynamics simulations using a coarse-grained RNA with explicit ions \citep{Denesyuk15NatChem}. In previous studies, we showed that the model reproduces \Mg{}-concentration dependence of the \Azo{} ribozyme folding and correctly predicts binding sites of \Mg{} in equilibrium simulations \citep{Denesyuk15NatChem, Hori19BJ}. Here, we focused on the kinetics of the same RNA. We conducted 95 folding simulations triggered by adding 5 mM \Mg{} to unfolded ribozyme prepared in the absence of divalent cations. Among them, 55 trajectories reached the native conformation within the simulation time. We found that a certain fraction of simulated trajectories were trapped in misfolded states. The folding reaction took place through multiple phases as monitored by the time-dependent changes in the overall size (\Rg{}), and formation of key interactions. Most ($\sim$80\%) of secondary structures folded rapidly within the first phase. In contrast, about half of tertiary interactions formed gradually during the first and the middle phases, and the other half folded in the last phase. Non-native base pairs contributed to a manifold of metastable states comprising of a combination of mispaired helices, which slowed the folding reaction. However, one of the misfolded states mostly consisted of native interactions without mispaired helices (a topological trap). Thus, not only non-native base pairs but also the topology of the chain are relevant in characterizing the rugged RNA folding landscape. We also analyzed the dynamics of \Mg{} ions, and showed that \Mg{} rapidly replaces \K{} when the folding reaction is initiated. Nearly 90\% of the number of \Mg{} ions were condensed onto the RNA in the earliest phase, in which most tertiary interactions and some helices were still unfolded. Comparison of our results, with several experimental data, including time-dependent \Rg{} from tSAXS experiments and hydroxyl radical footprinting, shows near quantitative agreement. This allows us to investigate the detailed structural changes triggered by \Mg{} as the \Azo{} ribozyme folds, events that cannot be accessed by ensemble or single molecule experiments. 

\section*{Materials and Methods}
\begin{figure*}
\includegraphics[width=0.65\textwidth]{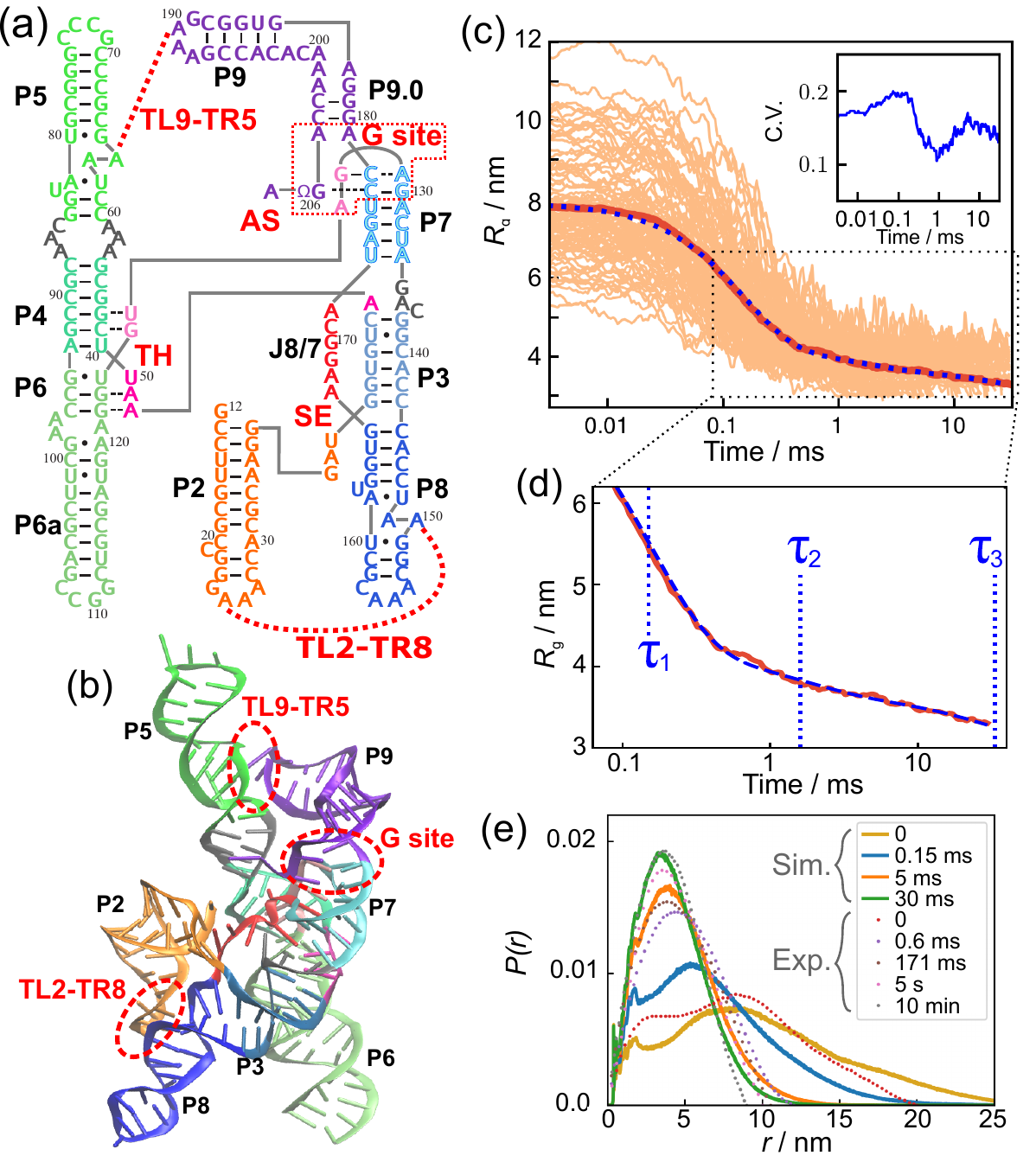}

\caption{
\label{fig:Fig1}
\textbf{Structure and \Mg{}-mediated collapse of the \Azo{} ribozyme.}
\textbf{(a)} The secondary structure map shows that several helices are ordered along the sequence. The helices of the \Azo{} ribozyme are conventionally denoted as P2 through P9 (labeled in black). P2, P5, P6, P8, and P9 are hairpin structures in which adjacent segments form the double strands, whereas P3, P4, and P7 are double strands formed by non-local pairs of segments. Several key elements involving tertiary interactions are shown in red: TH, triple helix; SE, stack exchange; G site, Guanosine-binding site; TL2-TR8, tetraloop 2 and tetraloop-receptor 8; TL9-TR5, tetraloop 9 and tetraloop-receptor 5; and AS, the active site.
\textbf{(b)} Tertiary structure, taken from PDB 1U6B \citep{AzoXray04N}. The same colors are used as in (a).
\textbf{(c)} Time dependence of the radius of gyration (\Rg{}) averaged over 95 trajectories (thick red line). Thin lines show individual trajectories. At the beginning of the simulations $t=0$, the average \Rg{} is $\meanRg{}\approx$~7.8~nm, corresponding to the value at equilibrium in the absence of \Mg{}.
The fit using three exponential functions (see the main text) is shown by the blue dotted line. Inset: Coefficient of variation (C.V.), calculated as $\sqrt{\left<\Rg{}(t)^2\right>-\left<\Rg{}(t)\right>^2}/\left<\Rg{}(t)\right>$ where $\left<\right>$ is the average over the trajectories.
\textbf{(d)} The middle to late stages of compaction are magnified from the data in (c). The three time constants are indicated by blue vertical lines. 
\textbf{(e)} Distance distribution functions reveal the stages in the collapse kinetics. The distribution functions were calculated using snapshots from the 95 trajectories at $t=0$, 0.15, 5, and 30 ms. The dotted lines are experimental tSAXS data from Ref.~\citep{Roh10JACS}.
}
\end{figure*}

\textbf{Three-Interaction-Site (TIS) model with Explicit ions:}
In order to simulate the long time scale needed to fold the ribozyme, we used the TIS model in the presence of \Mg{} as well as \K{} that is in the buffer \citep{Denesyuk15NatChem}. The TIS model for nucleic acids has three interaction sites for each nucleotide, corresponding to the phosphate, sugar, and base moiety (Fig.~S1) \citep{Hyeon05P}. All the ions are explicitly treated, whereas water is modeled implicitly using a temperature-dependent dielectric constant. The force field in which the physicochemical nature of RNA was carefully considered is given as $U_{\textrm{TIS}}=U_{\textrm{bond}}+U_{\textrm{angle}}+U_{\textrm{EV}}+U_{\textrm{HB}}+U_{\textrm{ST}}+U_{\textrm{ele}}$. The first two terms, $U_{\textrm{bond}}$ and $U_{\textrm{angle}}$ ensure the connectivity of the bases to the ribose backbone with appropriate bending rigidity. The next term, $U_{\textrm{EV}}$, accounts for excluded volume effects, which essentially prevent overlap between the beads. Hydrogen-bonding and stacking interactions are given by $U_{\textrm{HB}}$ and $U_{\textrm{ST}}$, respectively. We consider hydrogen bonds for all possible canonical pairs of bases (any G-C, A-U, or G-U base pairs can be formed), as well as tertiary hydrogen bonds that are formed in the crystal structure (PDB 1U6B). The stacking interactions are applied to any two bases from consecutive nucleotides along the sequence, as well as tertiary base stacking in the crystal structure. Parameters in these terms are optimized so that the model reproduces the thermodynamics of nucleotide dimers, several types of hairpins, and pseudoknots \citep{Denesyuk13JPCB}. It should be emphasized that {\it no parameter} in the energy function was adjusted to achieve agreement with experiments on the simulated ribozyme. Thus, the results are emergent consequences of direct simulations of the {\it transferable TIS model}. The detailed functional forms for these terms are given in the Supplementary Methods and Table S1 in the Supplementary Data. The optimized parameters and a list of tertiary interactions can be found in the main text and supplemental information given elsewhere \citep{Denesyuk15NatChem}. We showed previously that the model reproduced the experimental thermodynamics data for several RNA motifs, such as hairpin and pseudoknot, thermodynamics of \Azo{} ribozyme as a function of \Mg{} concentrations, and more recently, the thermodynamics of assembly of the central domain of the ribosomal RNA \cite{Hori21PNAS}. A crystal structure of \Azo{} group I intron (PDB 1U6B \citep{AzoXray04N}, Fig.~\ref{fig:Fig1}b) was used as the reference structure for the native conformation. The nucleotides are numbered from 12 through 207 following the convention in the literature of \Azo{} group I intron.

\medskip{}

\textbf{Simulation protocol:}
To generate the unfolded state ensemble, we performed equilibrium simulations in the absence of \Mg{} with 12~mM KCl, corresponding to the concentration in the Tris buffer. To enhance the efficiency of sampling the configurations of the system, we employed under-damped Langevin dynamics \cite{Honeycutt92} simulations by setting the friction coefficient to 1\% of water viscosity. We recorded the conformations of RNA and ions every $10^{8}$ steps, to be used as initial coordinates for the folding simulations. The duration of the equilibrium simulation is sufficiently long that all the initial structures are well separated in the configurational space. To generate 95 initial configurations, we ran over $3\times10^{11}$ steps constituting equilibrium simulations. Brownian dynamics simulations \citep{Ermak78JCP} were performed to trace the folding reactions starting from an initial state in which the ribozyme is devoid of tertiary structures. The viscosity was set to the value of water, $8.9\times10^{-4}\,\textrm{Pa\ensuremath{\cdot}s}$. The simulations were performed in a cubic 35~nm box containing ions and RNA. To minimize finite-size effects, we used periodic boundary conditions. The temperature was set to $T=37^{\circ}\textrm{C}$. Additional details are described in Supplementary Methods.

Starting from the initial conformations, prepared at [\Mg{}] = 0, we triggered the folding reaction by adding 5 mM \Mg{}. Both experiments and previous simulations \citep{Denesyuk15NatChem} have shown that a solution containing 5 mM \Mg{} and 12 mM \K{} drives \Azo{} ribozyme to structures that are catalytically active \citep{Roh10JACS}. We generated 95 folding trajectories until the ribozyme reaches the folded state, or the simulation time is $\approx 30\,\textrm{ms}$. We assessed whether the ribozyme is folded to the correct native state by calculating the root-mean-square-deviation (RMSD) from the crystal structure.
 If the RMSD to the native structure is less than 0.6 nm, then the ribozyme is folded. We confirmed that if RMSD < 0.6 nm, all the secondary and tertiary interactions are correctly formed. Note that the experimentally (tSAXS) accessible quantity, the radius of gyration (\Rg{}), alone is not an accurate order parameter to distinguish the native structure from the misfolded structures because some of the non-native structures have \Rg{} values that are close to the native structure. We use several measures, as described here and in the Supplementary Methods, to monitor the order of events during the folding process.

\medskip{}

\textbf{Clustering analysis:}
To classify the compact conformations of the ribozyme and examine if there are any non-native (misfolded) conformations, we performed a clustering analysis. First, from all conformations in the 95 folding trajectories, we collected compact structures whose $\Rg{}\leq$~3.5~nm, regardless of their similarities to the native conformation. This criterion generated 2,162,299 structures. After reducing the number of structures to 10,811 (1/200) by random selection, a clustering analysis was done by Ward's method using the Distribution of Reciprocal of Interatomic Distances (DRID) as the similarity measure \citep{DRID}.

\medskip{}

\textbf{Ion condensation and binding:} To make a quantitative comparison of condensation of monovalent (\K{}) and divalent (\Mg{}) cations, we counted the number of ions condensed onto the RNA at each time frame. We consider that an ion is condensed when it is in the vicinity of any phosphate site in the RNA. To compare \Mg{} and \K{} on equal footing, we used the Bjerrum length ($l_B = 0.73$ nm) as the cutoff distance for ion condensation. At the distance $l_B$ the thermal energy balances the Coulomb attraction between a cation and an anion.

To detect tightly-bound \Mg{} ions, in a manner consistent with the previous study \citep{Denesyuk15NatChem}, we computed the contact \Mg{} concentration, $c^{\ast}$, as follows. For every phosphate site in each snapshot of the simulations, we counted the number of \Mg{} located in the range, $r_0 - \Delta r < r < r_0 + \Delta r$, where $r$ is the distance from the phosphate, $r_0 = R_{\textrm{P}} + R_{\textrm{Mg}} = 0.44$ nm is the sum of the excluded-volume radii of the phosphate and \Mg{} ion, and $\Delta r = 0.15$ nm is a tolerance margin for contact (Fig.~S2a). The contact \Mg{} concentration was then calculated by dividing the number by the spherical shell volume (Fig.~S2). The definition of \Mg{} binding rate using the same criterion can be found in the Supplementary Methods.

\medskip{}

\textbf{Footprinting analysis:}
It is known that experimental footprinting using hydroxyl radical is highly correlated with the Solvent Accessible Surface Area (SASA) of the sugar backbone \citep{Cate96S,Balasubramanian98P,Ding12NatMeth}. To compare our simulation results with experimental footprinting data, we calculated SASA using FreeSASA version 2.0 \citep{FreeSASA16}. Considering that hydroxyl radicals preferably cleave C4' and C5' atoms of RNA backbone \citep{Balasubramanian98P}, we took the larger value of the SASA of C4' and C5' atoms for each nucleotide. From the SASA data, we computed the ``protection factor'' \citep{Adams04RNA} of the $i^{\textrm{th}}$ nucleotide (see the Supplementary Methods for the detail), $F_{p}^{\textrm{Native}}(i)=\frac{\left\langle SASA(i)\right\rangle _{\textrm{Unfolded}}}{\left\langle SASA(i)\right\rangle _{\textrm{Native}}}$ and $F_{p}^{\textrm{Misfold}}(i)=\frac{\left\langle SASA(i)\right\rangle _{\textrm{Unfolded}}}{\left\langle SASA(i)\right\rangle _{\textrm{Misfold}}}$ for the native and misfolded states, respectively, where the bracket indicates the ensemble average. We used the initial conformations prepared in the absence of \Mg{} to compute the average SASA of the unfolded state, $\left\langle \textrm{SASA}\right\rangle _{\textrm{Unfolded}}$. In order to perform the SASA calculation, we reconstructed atomically detailed structures of the ribozyme from coarse-grained coordinates using an in-house tool (Fig.~S1), which employs a fragment-assembly approach \citep{TIS2AA,Humphris12JMB}, followed by minimization by Sander in Amber16 \citep{Amber16}. In Fig.~S3, we show some examples of atomically detailed structures that were obtained from the conformations generated using the TIS model. 

\begin{figure*}
\includegraphics[width=0.8\textwidth]{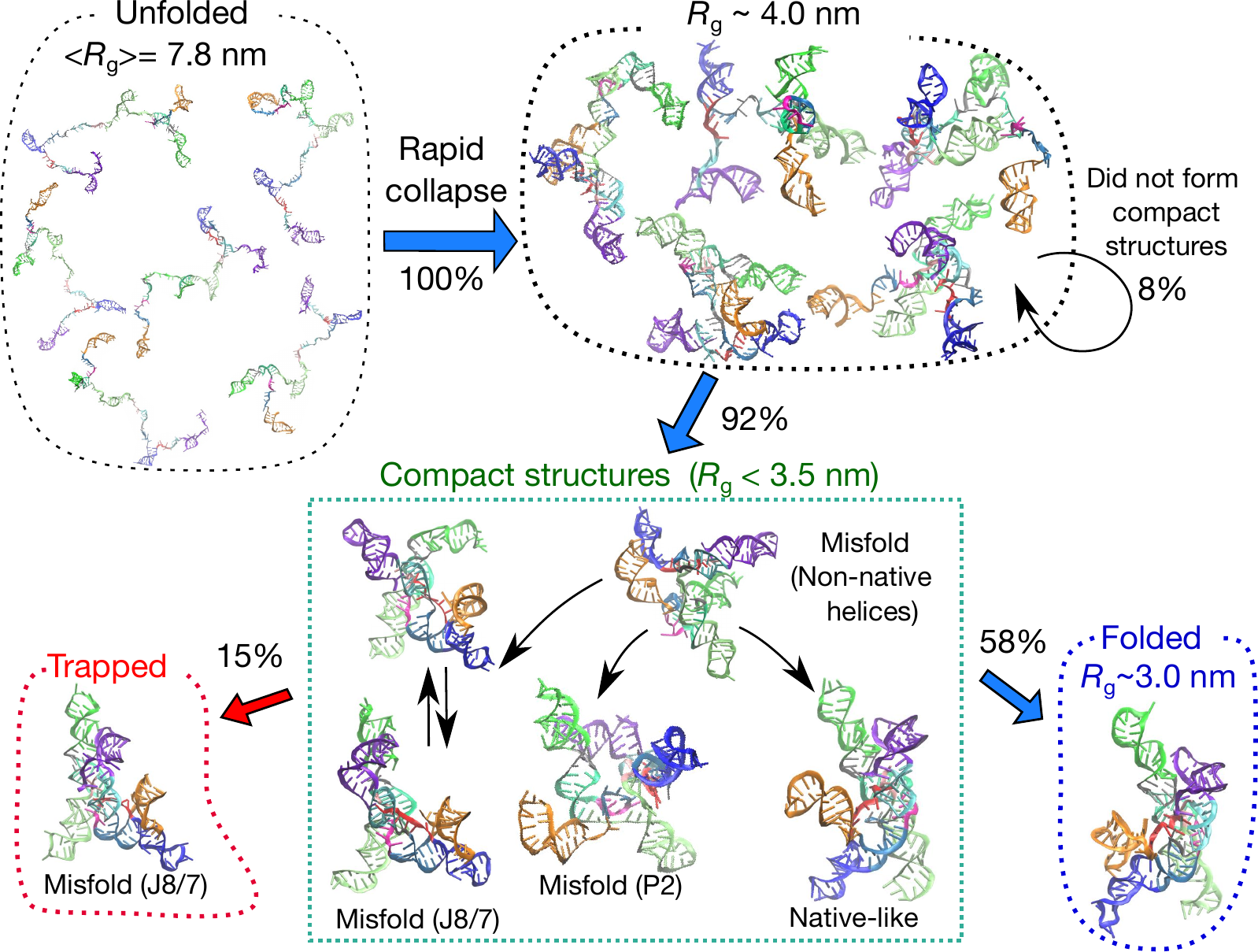}

\caption{\label{fig:Fig2}
\textbf{Kinetic partitioning of the trajectories.}
The blue and red arrows show the fate of the 95 folding trajectories initiated from the unfolded state (top left). The numbers beside the arrows indicate the fraction of trajectories in different pathways. In the compact structural ensemble, several representative misfolded structures, obtained from a clustering analysis, are shown. The structures labeled 'Misfold (P2)' (shown in the middle of the panel in the box) and 'Misfold (J8/7)' (lower left) are topological traps. Within the simulation time, 58\% of the trajectories reached the folded (native) state (right bottom), whereas 15\% were topologically trapped in the J8/7 misfolded state (left bottom). The remaining trajectories (27\%) are trapped in the compact-structure ensemble, partly because helices are mispaired (energetic trap).
}
\end{figure*}

\section*{Results}
There are two parts to the \Mg{}-induced folding of the ribozyme. One pertains to the time-dependent changes in the conformations of the RNA as it folds. The second is related to the role that \Mg{} plays in driving the ribozyme to the folded state. It is easier to infer the major conformational changes of the ribozyme using experiments. In contrast, it is currently almost impossible to monitor the fate of the time-dependent changes in many \Mg{} ions as they interact with specific sites on the RNA. Both the time-dependent changes in the RNA structures and, more importantly, how correlated motions involving multiple divalent cations lead to folding, cannot be simultaneously be measured in experiments. Simulations, provided they are reasonably accurate, are best suited to probe the finer details of RNA conformational changes, and the mechanism by which \Mg{} drives the ribozyme to fold as a function of time. After demonstrating that our simulations nearly quantitatively reproduce the measured time-dependent changes in the size of \Azo{} ribozyme, we focus on the effect of \Mg{} on the folding reaction.

\medskip{}

\textbf{Unfolded state ensemble is heterogeneous:}
We prepared the unfolded state structural ensemble in the absence of \Mg{} at 12 mM KCl concentration, which is the concentration in the Tris buffer \cite{Behrouzi12Cell}. The average radius of gyration, in the absence of \Mg{}, is $\meanRg{} = 7.8\,\textrm{nm}$ (Fig.~\ref{fig:Fig1}c), which is in excellent agreement with experiments ($\approx7.5\,\textrm{nm}$ measured by SAXS experiments in the limit of low [\Mg{}]) \citep{Behrouzi12Cell}. Although the tertiary interactions are fully disrupted, several secondary structures, helix domains P2, P4, P5, and P8, are almost intact (Fig.~S4). Nevertheless, globally the \Azo{} ribozyme is unstructured, with most of the characteristics of the native structure being absent (see the unfolded structures in Fig.~\ref{fig:Fig2}). The unfolded conformational ensemble is highly heterogeneous, containing a mixture of some secondary structural elements and flexible single-stranded regions (Fig.~S5).

\medskip{}

\textbf{Ribozyme collapse occurs in three stages:}
We first report how the folding reaction proceeds from the ensemble perspective by averaging over all the folding trajectories in order to compare with the tSAXS experiments \citep{Roh10JACS}. Following the experimental protocol, we monitored the folding kinetics using the time-dependent changes in the radius of gyration, \Rg{}, describing the overall compaction of the ribozyme (Fig.~\ref{fig:Fig1}c). At each time step, \Rg{} was averaged over all the trajectories. At $t\rightarrow0$ (see the limit of small $t$ in Fig.~\ref{fig:Fig1}c), the average is $\meanRg{} \approx7.8\,\textrm{nm}$, corresponding to the unfolded state. The time-dependent changes in the average \meanRg{} $\equiv R_g(t)$ are fit using,
\begin{equation}\label{eq:Rg(t)}
\Rg{}(t) = \Rg{}_\text{U} - (\Rg{}_\text{U} - \Rg{}_\text{F})
\sum_{i=1}^{3}{\Phi_{i}\left(1-e^{-\frac{t}{\tau_{i}}}\right)}
,
\end{equation}
where $\Rg{}_\textrm{U}$ and $\Rg{}_\textrm{F}$ are the average \meanRg{} of the unfolded and folded state, respectively, and $\tau_i$ and $\Phi_i$ are the time constants and the amplitudes ($\Phi_1 + \Phi_2 + \Phi_3 = 1$) associated with the $i$\textsuperscript{th} phase, respectively. The data could not be fit accurately using a sum of two exponential functions (Fig.~S6). The best fit parameters are $\Phi_1 = 0.76$, $\Phi_2 = 0.11$, $\Phi_3 = 0.13$, with the corresponding time constants, $\tau_1 = 0.15$, $\tau_2 = 1.6$, $\tau_3 = 33$ ms (Fig.~\ref{fig:Fig1}d). The three time scales describe the multi-step folding events if the radius of gyration is a reasonable order parameter for the folding reaction: (i) rapid collapse from the unfolded state ($\meanRg{} \approx7.8\,\textrm{nm}$) to an intermediate state in which the RNA is compact with $\meanRg{} \approx4\,\textrm{nm}$ ($\tau_c = \tau_1 = 0.15$ ms);
(ii) Further compaction to an intermediate state, $I_c$, which has $\meanRg{} \approx3.5\,\textrm{nm}$; and (iii) finally folding to the native structure with $\meanRg{} = 3\,\textrm{nm}$. Clearly, the maximum extent of compaction occurs in the earliest stage of the folding reaction.

It is important to compare the simulation results with experimental SAXS data \citep{Roh10JACS} in order to validate our model. The tSAXS results also show the three stages, with an initial decrease to $\meanRg{} \approx4\,\textrm{nm}$ occurring in $\tau_{1}^{\textrm{exp}} < 0.2$ ms. This is close to the time scale observed in the simulations, $\tau_1=0.15$ ms. In the tSAXS experiment, further compaction to the \Ic{} state ($\meanRg{} \approx3.5\,\textrm{nm}$) occurs in $\tau_{2}^{\textrm{exp}}\approx17\,\textrm{ms}$, which is an order of magnitude larger than predicted in the simulations, $\tau_2 = 1.6\,\textrm{ms}$. In the experiments \citep{Roh10JACS}, there is no data for \Rg{} for time less than $0.6$ ms, which might influence the estimate of  $\tau_{2}^{\textrm{exp}}$. The $\meanRg{}$ of the \Ic{} state is similar in the simulations and experiments (Fig.~\ref{fig:Fig1}d).

By fitting \Rg{} in the simulations to the three-stage kinetics, we obtained the time constant of the final transition leading to $\meanRg{} \approx3\,\textrm{nm}$, $\tau_3 = 33\,\textrm{ms}$, that is orders of magnitude smaller than the experimental estimate  ($\tau_{3}^{\textrm{exp}} \approx 170\,\textrm{s}$). The most likely reason for this discrepancy is that, for computational reasons, we terminated the folding trajectories at 30 ms regardless of whether the RNA is folded or not. Therefore,  the estimated value  of $\tau_3$ from simulations  is a lower bound to the time constant reported in experiments. It is worth noting  that analysis of the experimental data emphasized only $\tau_{1}^{\textrm{exp}}$ and $\tau_{2}^{\textrm{exp}}$, spanning times on the order of $200\,\textrm{ms}$ (see Fig. 4A in \citep{Roh10JACS}).  

The trajectories that do not reach the native state in 30 ms undergo non-specific collapse into structures that require a longer time to reach the folded state. Nevertheless, it is clear that the simulations capture the multistage collapse of the ribozyme observed in tSAXS experiments, with near quantitative agreement for the first two phases (Fig.~\ref{fig:Fig1}d). We confirmed that distance distribution functions at several different stages in the folding are also consistent with experiments \citep{Roh10JACS} (Fig.~\ref{fig:Fig1}e).

The amplitudes of all the three stages in the decay of $\Rg{}(t)$ in the simulations are in a near quantitative agreement with fits to the measured $\Rg{}(t)$ (see Figure 4 in \cite{Roh10JACS}). At [\Mg{}] = 5 mM, the calculated values from the simulations are $\Phi_1 = 0.76$, $\Phi_2 = 0.11$, and $\Phi_3=0.13$, whereas the experimental estimates are $\Phi_{1}^{\textrm{exp}} = 0.8$, $\Phi_{2}^{\textrm{exp}} = 0.1$, and $\Phi_{3}^{\textrm{exp}} = 0.1$. The level of agreement is remarkable, considering that no parameter in the model was adjusted to obtain agreement with any observable in the experiment. If $R_g$ is a good reaction coordinate, then this would imply that nearly 80\% of \Azo{} ribozyme folds rapidly.

The ensemble average \meanRg{} hides the high degree of heterogeneity in the \Mg{}-induced compaction of the RNA. The results in the inset of Fig.~\ref{fig:Fig1}c show that the dispersion in \Rg{} as a function of $t$ is considerable even in the late stages of folding. This is the first indication of the importance of pathway heterogeneity in the folding process, which we further substantiate below. 

\medskip{}

\textbf{Dynamics of folding and misfolding:}
In Fig.~\ref{fig:Fig2}, the fate of the trajectories and schematic folding routes are presented along with some representative structures. After the initial compaction, most trajectories (87 out of 95 $\approx$ 92\%) show a further reduction in \Rg{} in the second phase, in which compact structures form ($R_{g}<3.5$ nm). Among them, 55 trajectories reach the folded state in 30 ms with structural features that are the same as in the crystal structure. To identify the distinct conformations that appear in the compact structural ensemble, we performed a clustering analysis (Supplementary Methods and Fig.~S7). Although there are several misfolded elements in the ensemble structures, along with native-like structures, we identified two types of misfolding mechanisms, namely energetic traps and topological traps. The energetic trap consists of misfolded states that contain non-native helices (Fig.~S8). These non-native helices can be eventually disrupted (two strands dissociate by stochastic thermal fluctuations), and undergo transition either to the native-like state or to one of the second type of misfolded states.

The second type of misfolded conformations, which we classify as topological traps, are due to frustration arising from chain connectivity \citep{Guo95BP}. In the topological trap, most of the helices are correctly formed, as in the native structure. However, the spatial arrangement of helices in some regions differs from the native structure (e.g., a part of the chain passes either on one side or the other of another strand). The incorrect topology is stabilized by several tertiary interactions, especially those involving peripheral motifs such as TL2-TR8 and TL9-TR5 (Fig.~\ref{fig:Fig1}). As a consequence, once RNA is topologically trapped, it cannot easily escape and fold to the correct native state at any reasonable time.

We identified two distinct topological traps. In the major topological trap, `Misfold (J8/7)', the J8/7 junction passes through an incorrect relative location with respect to the strands of the P3 helix (Fig.~\ref{fig:Fig7}b). We describe this major topological trap in detail in the following sections. Although `Misfold (J8/7)' resembles the native structure, the lifetime of this state is so long that it cannot be resolved even on the experimental time scale. In the minor topological trap labeled as `Misfold (P2)', a part of the chain leading to the P2 helix is incorrect. This structure is stabilized by TL9-TR5 peripheral tertiary contact (See Supplementary Discussions and Figs.~S9--S11).

In summary, within the simulation time of 30 ms, 55 out of 95 trajectories folded correctly to the native structure. In 14 trajectories, the ribozyme was trapped in the misfolded (J8/7) state, which is a native-like topological trap. In 18 out of the remaining 26 trajectories, compact structures formed ($\Rg{}<3.5$ nm) rapidly but did not fold further, partly due to the energetic or topological frustration. Because folding is a stochastic process, the times at which each trajectory (or molecule) reaches the native state vary greatly.

\begin{figure*}
\includegraphics[width=0.7\textwidth]{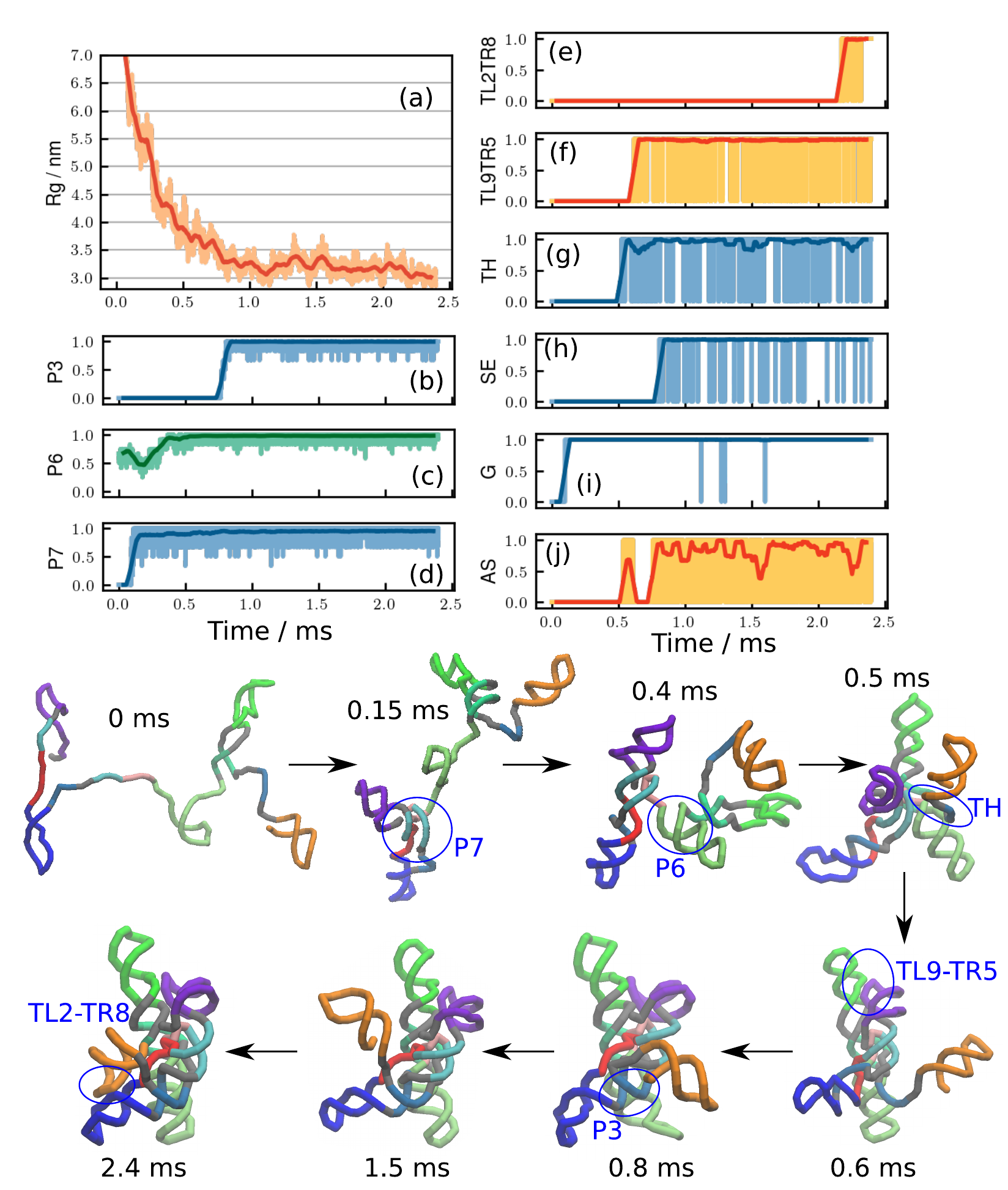}

\caption{
\label{fig:Fig3}
\textbf{A representative rapid folding trajectory.}
The ribozyme folds in $t\sim2.4\,\textrm{ms}$ without being kinetically trapped.
\textbf{(a)} Time dependent changes in $R_{g}$.
\textbf{(b-d)} Fraction of helix formations for (b) P3, (c) P6, and (d)P7.
\textbf{(e-j)} Fraction of major tertiary interactions (e) TL2-TR8, (f) TL9-TR5, (g) Triple Helix, (h) Stack Exchange, (i) G site, and (j) Active Site. See Fig.~\ref{fig:Fig1}(a-b) for the locations of these structural elements. Thin lines, with light colors are raw data, and thick lines with dark colors are averaged over 50 $\mu$s window.
\textbf{(Bottom)} Eight representative structures at several different time points. Major conformational changes are indicated by blue circles with labels. See Supplementary Movie 1 to watch the trajectory.
}
\end{figure*}

{\bf Visualizing folding events in a single folding trajectory:} In Fig.~\ref{fig:Fig3}, we show a representative trajectory in which folding is completed (see also Supplementary Movie 1). In this particular case, the ribozyme reached the native structure rapidly in $t\sim2.4\,\textrm{ms}$. Helices P2, P4, P5, P8, and P9 were already formed at $t=0$, and remained intact during the folding process. From $t=0$ to $\sim0.5\,\textrm{ms}$, global collapse occurred, with a rapid decrease in \Rg{} to $\sim$4 nm. Along with the global collapse, P7 formed at an early stage, $t\sim0.15$ ms. Helix P6 and the G-site tertiary interaction formed around $t\sim0.4$ ms. Other key interactions formed in the order of Triple Helix (TH), TL9-TR5, and P3. Note that the order of formations of these key interactions depends on the trajectory, and is by no means unique. After the formation of the P3 helix, it took a relatively long time ($\sim1.5$ ms) for P2 (orange domain in Fig.~\ref{fig:Fig3}) to find the counterpart P8 (blue). At $t\sim2.4$ ms, the formation of tertiary interactions between the two domains (TL2-TR8) results in the native conformation. Figs.~S12 and S13 show two other examples in which the folding was completed but on a longer time scale. In the trajectory in Fig.~S12, a mispaired helix in P6 formed early ($t<1$ ms), preventing it from folding to the native state smoothly. Eventually ($t\sim25$ ms), the misfolded P6 was resolved, leading to the correct native fold.

\begin{figure*}
\includegraphics[width=0.6\textwidth]{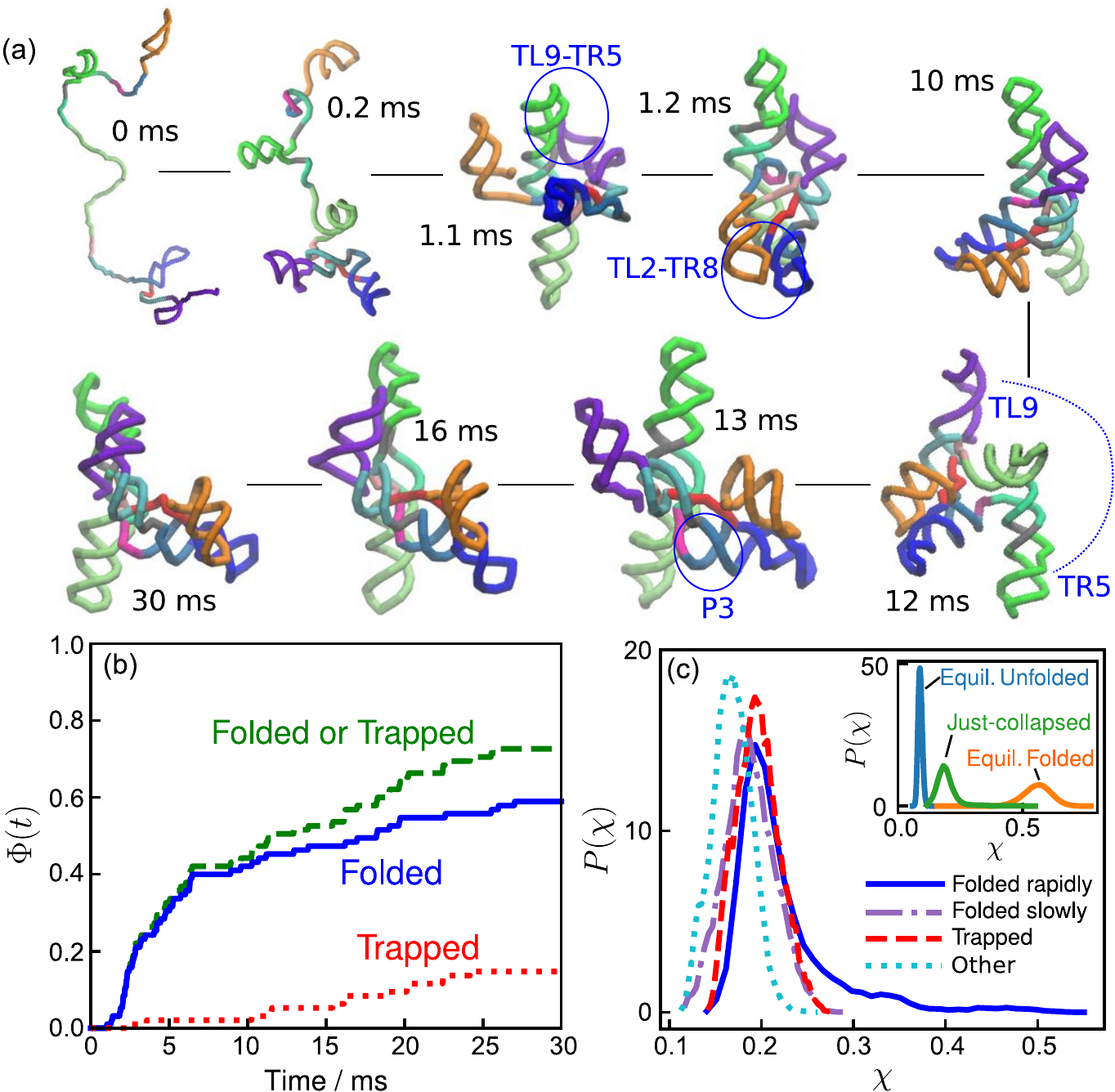}

\caption{
\label{fig:Fig4}
\textbf{Propensity to fold correctly is determined early.}
\textbf{(a)} A representative misfolding trajectory showing the formation of long-lived topologically-trapped states. Two key interactions in the peripheral regions, TL2-TR8 and TL9-TR5, formed early in the folding process ($t < 1$ ms), resulting in the junction J8/7 with an incorrect topology (See Fig.~\ref{fig:Fig7}). Because the incorrect chain topology cannot be resolved unless both of the peripheral interactions unfold, the RNA stays topologically trapped for the rest of the simulation time. See Fig.~S14 for trajectories of \Rg{}, fractions of secondary and tertiary elements, Fig.~S15 for another example of a kinetic trap, and Supplementary Movie 2 for watching the trajectory.
\textbf{(b)} Specific collapse leads to rapid folding. From the distributions of the first passage times to the folded state, $P_{FP}^{F}(\tau_{F})$, we calculated,
$\Phi^F(t) = \int_{0}^{t}{P_{FP}^{F}(s)ds}$.
Similarly,
$\Phi^M(t) = \int_{0}^{t}{P_{FP}^{M}(s)ds}$,
where $P_{FP}^{M}$ is the distribution of mean first passage times to the trapped state.
\textbf{(c)} Fate of the ribozyme immediately following the initial collapse. The probability distributions of the structure overlap ($\chi$) with respect to the native structure; $\chi=0$ indicates no similarity to the crystal structure, and $\chi=1$ corresponds to the native state. (Inset) The distribution of $\chi$ immediately after collapse ($t < 150$ $\mu$s, green line, "Just-collapsed") compared with distributions of the equilibrium unfolded state (blue, [\Mg] = 0 mM) and folded state (orange, [\Mg] = 5 mM). The distribution of the "just-collapsed" ensemble in the main figure is decomposed into four distributions depending on the fate of each trajectory: Folded rapidly, trajectories reached the correct folded state within 5 ms; Folded slowly, trajectories reached the correct folded state after 5 ms but within the maximum simulation time (30 ms); Trapped, trajectories where the RNA was trapped in the major misfolded state; trajectories labeled Other were neither folded nor misfolded.
}
\end{figure*}

Fig.~\ref{fig:Fig4}(a) shows a series of snapshots from a trajectory where the ribozyme is topologically trapped in the Misfold (J8/7) state. In this trajectory, two key interactions in the peripheral regions, TL2-TR8 and TL9-TR5, formed early ($t < 1$ ms). However, the junction J8/7 was in the wrong position with respect to the P3 helix strands, causing a misfolding (Fig.~\ref{fig:Fig7}b). Because the incorrect chain topology cannot be resolved unless both of the peripheral interactions unfold, the RNA stays in this misfolded topological trap for an arbitrarily long time.

From all the folding and misfolding trajectories, we calculated the distributions of first passage times to either the folded or the trapped state. The time-dependent fractions in these two states, shown in Fig.~\ref{fig:Fig4}(b), reveal that trajectories that reach native-like structures (\textit{i.e.} either folded or topologically trapped states) by $\sim$5 ms fold correctly. In contrast, trajectories that take a longer time to be native-like ($>$ 5 ms) are kinetically trapped. The former type of trajectories would be the consequence of the specific collapse in the earliest stage of the folding.

\medskip{}

\begin{figure*}
\includegraphics[width=0.9\textwidth]{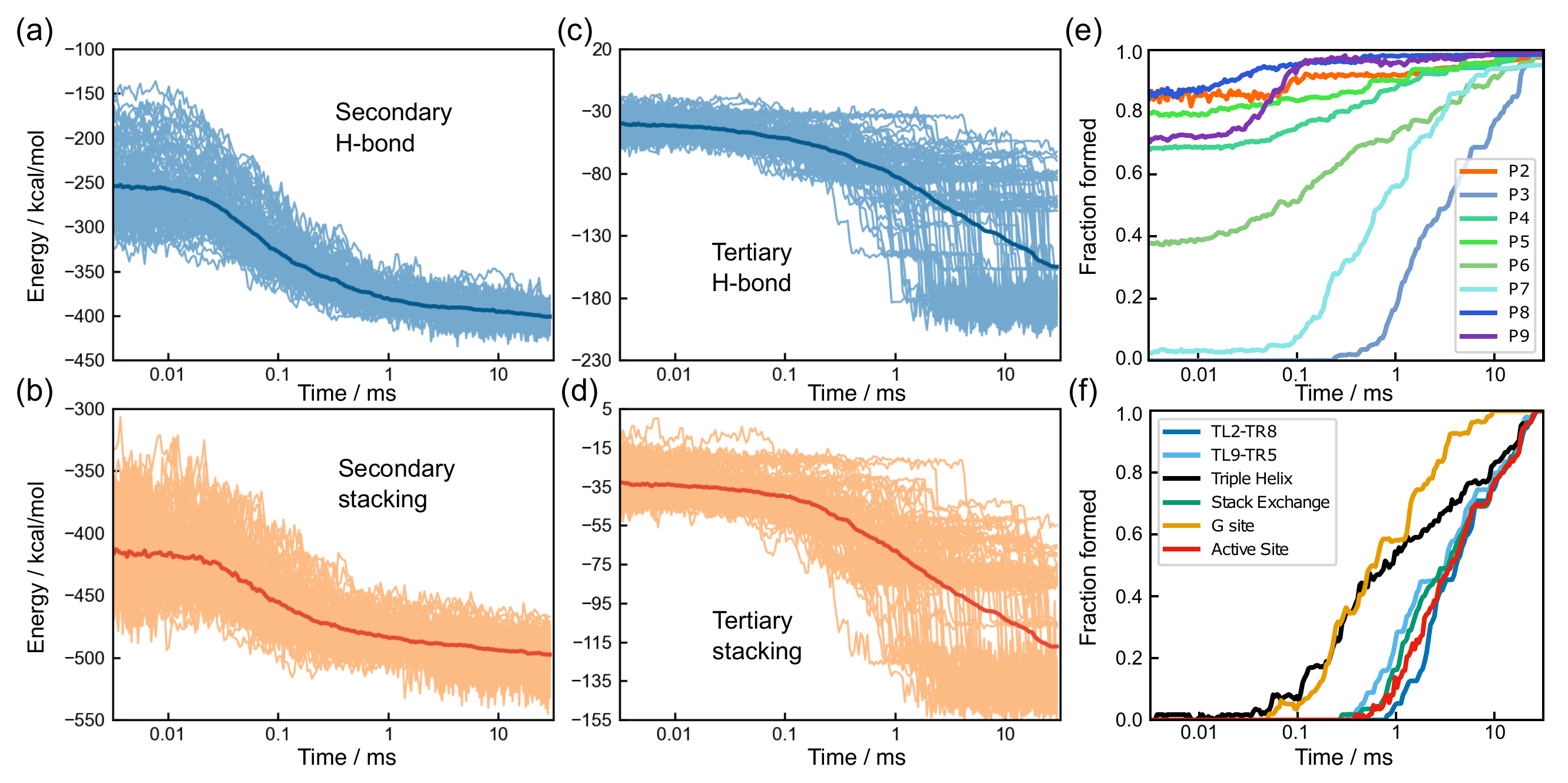}

\caption{
\label{fig:Fig5}
\textbf{Hierarchical formation of secondary and tertiary structures.} (a-d) Time-dependent formation of structural elements represented by the potential energies associated with (a) secondary hydrogen bond (H-bond), (b) secondary stacking, (c) tertiary H-bond, and (d) tertiary stacking.
The thin lines are the results for the 95 individual trajectories, and the thick line in each panel is the average over all trajectories.
(e) Time dependence of helix formation for helices P2 to P9.
(f) Formation of six key elements associated with tertiary interactions (see Fig.~\ref{fig:Fig1}).
}
\end{figure*}

\textbf{Kinetics of secondary- and tertiary-structure formation:}
The folded RNA structures are composed of secondary structural motifs, which are often independently stable and are consolidated by tertiary contacts to render the ribozyme compact. The most abundant elementary structural unit is the double-stranded helix (Fig.~\ref{fig:Fig1}a). In Fig.~\ref{fig:Fig5}, time-dependent formations of secondary and tertiary interactions, represented by average energies stabilizing these motifs, are plotted. Each interaction type is further categorized into two main chemical components, hydrogen bonding (H-bond) and base stacking. Fig.~\ref{fig:Fig5} (a-b) shows that most secondary structures are rapidly formed in the first phase ($t\lesssim0.15\,\textrm{ms}$), although certain secondary interactions form only in the late stages. In contrast, the formation of most tertiary interactions occurs in the middle ($t\sim1.5$ ms) and the last phase ($t\gtrsim10\,\textrm{ms}$) (Fig.~\ref{fig:Fig5}c-d). These findings illustrate the hierarchical nature of RNA folding kinetics in the ensemble picture, where formations of secondary structures are followed by tertiary contacts \cite{Brion97ARBiophys,Tinoco99JMB}.

We then investigated the kinetics of individual helix formations by calculating time-dependent fractions of helix formations in the folding trajectories (Fig.~\ref{fig:Fig5}e). In summary, all helices except P3 and P7 fold early, typically in $t<0.1$ ms. Kinetics of involving helices P3 and P7 are particularly slow because the two strands of P3 and P7 are far apart in the sequence, and thus it takes substantial time to search each other. Supplementary Discussion contains further details on the folding of individual secondary structures.

Equilibrium ensemble simulations \citep{Denesyuk15NatChem} identified several key tertiary interactions (shown in red symbols in Fig.~\ref{fig:Fig1}) by varying the \Mg{} concentration. In Fig.~\ref{fig:Fig5}(f), formations of those tertiary interactions are shown as time averages over the folded trajectories. Here, we find that the triple helix (TH) and Guanosine binding site (G site) form earlier than the other key elements. This is consistent with the results of equilibrium simulations \citep{Denesyuk15NatChem} that reported that TH and G sites are formed at lower \Mg{} concentrations compared to other key interactions. The time range of the formation corresponds to the second phase in Fig.~\ref{fig:Fig1}(c). Following the TH and G site formation, other key interactions form, mainly in the last phase ($t>10\,\textrm{ms}$). 

Interestingly, the kinetics of G site formation resembles the formation kinetics of the P7 helix (Fig.~\ref{fig:Fig5}e). From the secondary structure (Fig.~\ref{fig:Fig1}), this can be explained by noting that the G site is formed with a part of the P7 helix. Since the two strands of P7 are separated along the sequence, the formation of the P7 helix is a rate-limiting step for the formation of the G site. This is reminiscent of the diffusion-collision model proposed for protein folding \cite{Karplus76Nature}.
\medskip{}

\begin{figure*}
\includegraphics[width=\textwidth]{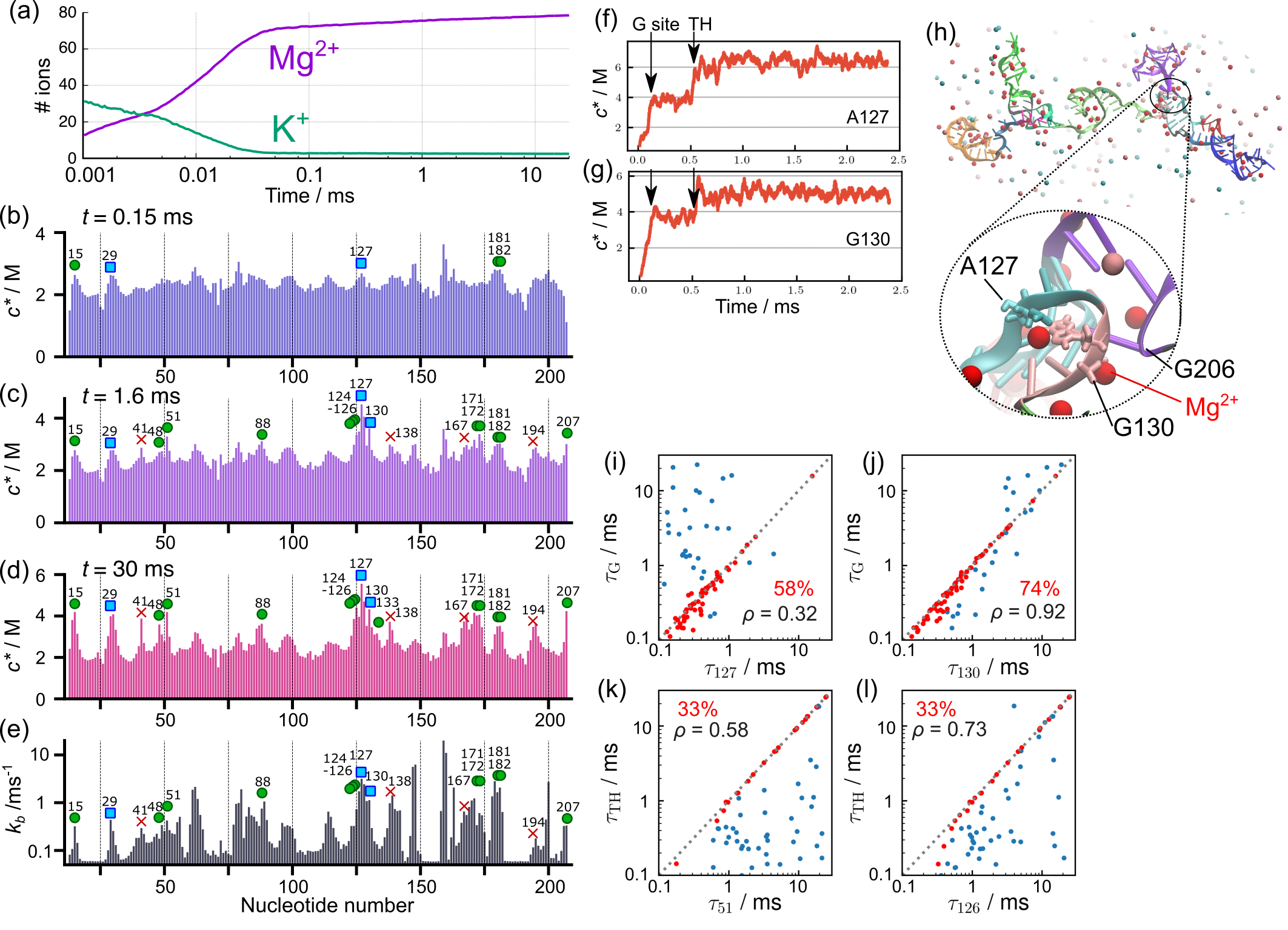}

\caption{
\label{fig:Fig6}
\textbf{Fingerprints of \Mg{} association and correlation with tertiary contact formation.}
\textbf{(a)} Number of cations condensed onto the RNA averaged over all the trajectories.
At $t=0$, on average, there are 40 \K{} ions in the neighborhood of the ribozyme. \K{} cations are rapidly replaced by \Mg{} in $t\apprle0.02\,\textrm{ms}$.
\textbf{(b-d)} \Mg{} fingerprints, measured using local nucleotide-specific concentration at (b) $t=0.15$ ms, (c) 1.6 ms, and (d) 30 ms. Symbols indicate nucleotides that are either in direct contact with \Mg{} (green circles) or linked via a water molecule (cyan squares) in the crystal structure \citep{AzoXray04N}. The red crosses are nucleotides predicted to bind \Mg{} in the equilibrium simulations \citep{Denesyuk15NatChem}.
\textbf{(e)} Binding rates calculated from the mean first passage times of \Mg{} binding event at each nucleotide.
\textbf{(f, g)} Trajectories displaying contact \Mg{} concentration at nucleotides (f) A127 and (g) G130 taken from the folding trajectory shown in Fig.~\ref{fig:Fig3}. The folding times of the G site and Triple Helix (TH) are indicated by black arrows. 
\textbf{(h)} A snapshot, at $t=0.11$ ms, when the G site is formed in the trajectory in (f) and (g). Nucleotides A127 and G130 are shown in stick, and \Mg{} ions are spheres in red.
\textbf{(i, j)} Scatter plots of the first passage times of \Mg{} binding to (i) A127 and (j) G130 versus the first passage time of the G site formation. If the \Mg{} binding and the G site formation occurred concurrently ($|\tau_{\textrm{G}} - \tau_i| < 0.2$ ms), the two events are regarded as strongly correlated and plotted in red; otherwise it is shown in blue. In each panel, the fraction of strong correlations (red points) is indicated by the percentage in red, associated with the Pearson correlation ($\rho$) calculated using all the data points.
\textbf{(k, l)} Same as (i-j) except for \Mg{} binding to (k) U51 and (l) U126 versus the first passage time of Triple Helix (TH) formation. 
}
\end{figure*}

\begin{figure*}
\includegraphics[scale=0.7]{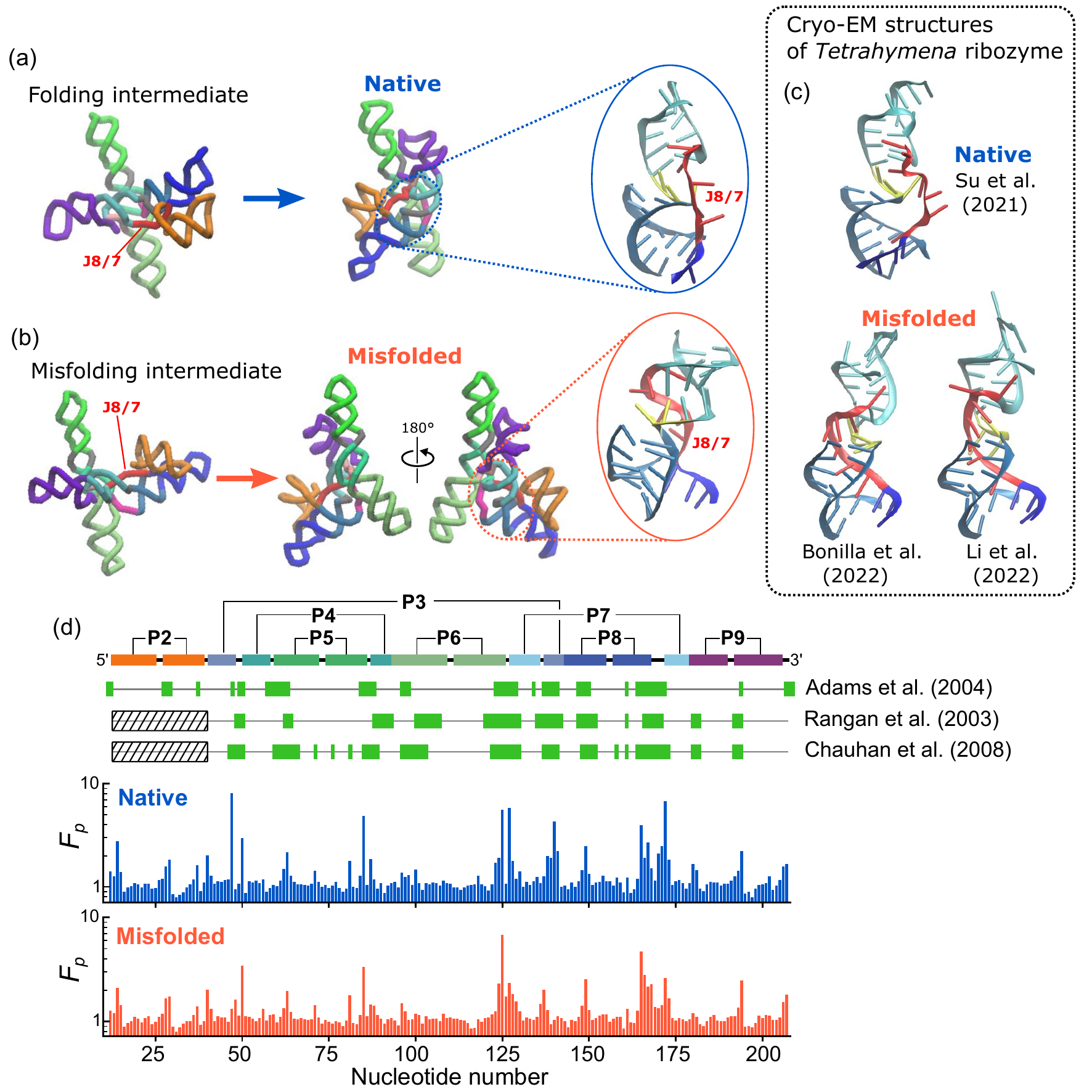}

\caption{
\label{fig:Fig7}
\textbf{Topological frustration in the persistent metastable state.}
\textbf{(a-c)} The major misfolded structure (J8/7) and its intermediate (b) are compared with the native intermediate and the folded structures (a). The spatial arrangements of J8/7 and other strands at the core are depicted on the right in blue and red circles. In the misfolding intermediate (b, left), the strand J8/7 (colored in red) passes through an incorrect location relative to the other two strands of the P3 helix (cyan), resulting in a topologically trapped state. This incorrect topology cannot be resolved unless other tertiary contacts such as TL2-TR8 (contact between domains P2 (orange) and P8 (blue)) disengage, which is unlikely to occur under the folding condition. As a consequence, it leads to a metastable state that is as compact as the folded state but with an incorrect chain topology. In panel (c), the same region in the native (upper, PDB 7EZ0 \cite{Su21Nature}) and misfolded (lower, PDB 7UVT \cite{Bonilla22SciAdv} and PDB 7XSK \cite{Li22PNAS}) \Tet{} ribozyme, solved by cryo-EM are shown for comparison. 
\textbf{(d)} Comparison of experimental footprinting data with SASAs calculated from simulations. Protection factor ($F_{p}$) are calculated for the native (middle, blue) and misfolded ensembles (bottom, red) from simulations. Protected nucleotides, indicated by green-filled rectangles on top, are from data in three experiments, as labeled on the right \citep{Adams04RNA,Rangan03PNAS,Chauhan08JACS}. The secondary structure is shown on top for reference. Note that protections of nucleotides in the P2 helix were not resolved in two experiments for technical reasons.
}
\end{figure*}

\textbf{Counterion release kinetics:}
We now turn to the role of the cations, \K{} and \Mg{}, in driving the assembly of \Azo{} ribozyme. The interplay between the unbinding of \K{} ions and the association of \Mg{} to the ribozyme as it folds is vividly illustrated in Fig.~\ref{fig:Fig6}(a). The figure shows the time dependence of the number of cations condensed onto RNA, averaged over all the trajectories (see Methods for the definition). 
 At $t=0$, on average, 40 \K{} are condensed onto the ribozyme (see Fig.~S16 for snapshots). The monovalent \K{} cations are rapidly replaced by \Mg{} in $t\apprle0.02\,\textrm{ms}$, which shows dramatically the counterion release mechanism anticipated by the application \citep{HeilmanMillerI01JMB, Hori19BJ} of the Oosawa-Manning theory \citep{Oosawa71Book}. In this time scale, nearly 90\% of \Mg{} ions in the final native state are already condensed, even though most of the tertiary interactions and some helices are still unfolded. The replacement of \K{} ions by \Mg{} shows that even the initiation of folding requires the reduction in the effective charge on the phosphate groups, which is accomplished efficiently by divalent cations. However, in this rapid process, some of the structures that form are topologically or energetically frustrated, thus greatly increasing the folding time. The premature condensation of \Mg{} has been found to produce kinetically heterogeneous structures that rearrange slowly both in Holliday junctions \cite{Hyeon12NatChem}, and group II introns \cite{Kowerko15PNAS}. A fraction of unfolded RNAs undergoes specific collapse, adopting compact structures that reach the native-like fold rapidly.

\medskip{}

\textbf{Fingerprint of \Mg{} associations:}
Among the condensed \Mg{} ions, some bind at specific sites as seen in the crystal structure \citep{AzoXray04N}. In Fig.~\ref{fig:Fig6}(b-d), we show the time-dependent \Mg{} densities, $c^{\ast}$, at each nucleotide site.
At many of the specific binding sites, \Mg{} ions associate with the ribozyme in the early stage of folding. For instance, at $t=0.15$ ms, although most tertiary interactions are not formed (Fig.~\ref{fig:Fig5}f), there are several peaks in the \Mg{} densities (Fig.~\ref{fig:Fig6}b). Interestingly, nucleotides 15, 29, 127, and 181 are the \Mg{} binding sites in the crystal structure. This shows that some of the specific binding sites are occupied by \Mg{} at the earliest stage of folding. Coordination of \Mg{} at these sites is required to reduce the electrostatic penalty for subsequent tertiary structure formation. In the intermediate time scale, $t=1.6$ ms, additional nucleotides bind \Mg{}, as shown in Fig.~\ref{fig:Fig6}(c). At $t=30$ ms (Fig.~\ref{fig:Fig6}d), we confirmed that \Mg{} binding sites are consistent with the crystal structure \citep{AzoXray04N}, which is another indication that the model is accurate.

We next investigated if there is a correlation between \Mg{}-binding kinetics and the thermodynamics of ion association. Using the same distance criterion for \Mg{} binding to the RNA, we computed the mean first passage time (MFPT) of \Mg{} coordination to each phosphate site. The binding rate was calculated as the inverse of the MFPT and is shown in Fig.~\ref{fig:Fig6}(e). The nucleotides that have greater binding rates ($k_b$) correspond to those that have higher \Mg{} densities in the equilibrium simulations \citep{Denesyuk15NatChem}, including the positions found in the crystal structure \citep{AzoXray04N}. This shows that the association of the ions to high-density \Mg{} sites occurs rapidly in the early stage of the folding.
These results show that there is a remarkable consistency between the order of accumulation of \Mg{} ions at specific sites of the ribozyme and the conclusions reached based on equilibrium titration involving an increase in \Mg{} concentrations. The coordination of \Mg{} at early times to nucleotides are also the ones to which \Mg{} ions bind at the lowest \Mg{} concentration \citep{Denesyuk15NatChem}, thus linking the thermodynamics and kinetics of ion association to the ribozyme.

\medskip{}

\textbf{Specific \Mg{} binding drives formation of tertiary interactions:}
From the kinetic simulation trajectories, we further analyzed whether \Mg{} binding events directly guide the formation of tertiary interactions.
One of the key tertiary elements that folds in the early stage is the G site comprising the G-binding pocket and G206 (Fig.~\ref{fig:Fig1}). There are two peaks corresponding to A127 and G130 in the \Mg{} fingerprint both in kinetics (Fig.~\ref{fig:Fig6}e) and thermodynamics \citep{Denesyuk15NatChem} simulations, consistent with a \Mg{} ion bound between the two nucleotides in the crystal structure \citep{AzoXray04N}. We found a strong correlation between the first passage times of \Mg{} binding at these nucleotides, and the formation of the G site. For example, in the same trajectory, as shown in Fig.~\ref{fig:Fig3}, \Mg{} binding to those nucleotides are noticeable as an increase in contact \Mg{} concentration ($c^{\ast}$) at the same time as G site formation (Fig.~\ref{fig:Fig6} f-g). 

The scatter plots in Fig.~\ref{fig:Fig6}(i, j) show the correlation between the first passage time for \Mg{} binding ($\tau_{127}$ and $\tau_{130}$) and the first passage time for G-site formation ($\tau_{\textrm{G}}$) from all the trajectories. For both A127 and G130, there is a distinct temporal correlation, reflected as dense data points in the diagonal region. In the majority of the trajectories (74\% for G130 and 58\% for A127), the G-site formation and \Mg{} binding occurred simultaneously within 0.2 ms (shown as red points in Fig~\ref{fig:Fig6}).
Overall, \Mg{} binding to G130 exhibited a higher correlation (Pearson coefficient $\rho = 0.92$), whereas the correlation for A127 was lower ($\rho = 0.32$) because G-site formation occurred later than the \Mg{} binding in some trajectories (data points on the upper left triangle in Fig.~\ref{fig:Fig6}i). Nevertheless, for both the nucleotides, the G-site formation does not precede \Mg{} binding, indicated by the absence of data points on the lower right triangle. We conclude that \Mg{} binding to these nucleotides is a necessary condition for the formation of the G site.

Strikingly, there are also noticeable increases in the local \Mg{} concentration at nucleotides A127 and G130 in the later stage, corresponding to the time when another tertiary element, Triple Helix (TH), forms (also indicated by arrows in Fig.~\ref{fig:Fig6}(f-g)). Because the G site and TH are spatially close to each other, the formation of TH further stabilizes the \Mg{} associations at the G site, especially A127, which is located at the end of the strand that constitutes the TH. The temporal correlations between these nucleotides and both the G site and TH show that \Mg{} binding to some nucleotides in the core of the ribozyme contribute to more than one tertiary contact.

Fig.~\ref{fig:Fig6} (k, l) show that U51 and U126 also bind \Mg{} ions upon folding of TH, which is consistent with the observation of two \Mg{} ions in the crystal structure. The scatter plots show that, in some trajectories, TH forms before \Mg{} ions bind to U51 and U126, indicating that \Mg{} binding to these nucleotides are not needed for the formation of the TH. This supports the finding that TH formed in 60\% of the population even at submillimolar \Mg{} concentration in the equilibrium simulation study \citep{Denesyuk15NatChem}. We found similar correlations for other key tertiary elements, Stack Exchange, TL2-TR8, and TL9-TR5. See Supplementary Discussion and Figs.~S17--S20.

\medskip{}

\textbf{Non-native base pairs impede folding both to the native and topologically-trapped states:}
Because our model allows any combination of canonical (G-C and A-U) and Wobble (G-U) base pairs to form, we found a number of non-native base pairs in the folding process. Some of these base pairs formed only transiently, whereas others have relatively long lifetimes, suggesting a link between the formation of non-native base pairs and misfolded states.
By counting the frequencies of each non-native base pair, we identified five frequently mispaired strand-strand combinations (Fig.~S8). These strands are parts of helices, P3, P6, P7, and junction J8/7. For instance, each strand of P3 forms a double-strand helix with another strand from P7, leading to the formation of mispaired helices P3u-P7u and P3d-P7d (Fig.~S8 right top). Here, we introduced the notation, \textit{u} and \textit{d}, to distinguish between the two strands for native helices, the upstream (5'-end) strand \textit{u} and the one downstream \textit{d}. Helices, P3, P6, P7 are unfolded in the absence of \Mg{} (Fig.~S4). Given that these mispaired helices consist of at least four consecutive base pairs, except P3d-J8/7, it is reasonable that such incorrect pairings are formed in the process of the unfolded strands searching for their counterparts. In addition, we found alternative secondary structures of P6 helix, alt-P6 (Fig.~S8 bottom right).

In Fig.~S21(a), we show the time-dependent fractions of mispaired helices. One of such mispairing, P3d-J8/7, formed earlier than others, and the fraction is also higher. Interestingly, there is a significant fraction of such mispaired helices even in folding trajectories that reach the native state. Indeed, the time-dependent fraction for folded and misfolded trajectories resemble each other (Fig.~S21 b-c). In both cases, the fractions of mispaired helices increase between $0.1<t<1$ ms, and then decrease to nearly zero at $t\sim30$ ms. In contrast to expectation, the persistent misfolded states do not have a significant amount of mispaired helices. We conclude that the mispaired helices (energetics traps) often form regardless of whether folding is on the pathway to the native state or kinetically trapped. 

\medskip{}

\textbf{Footprinting data is consistent with the formation of misfolded states:}
Hydroxyl radicals are often used to study hierarchical structure formations in footprinting experiments. In hydroxyl radical footprinting (analogous to hydrogen exchange experiments using NMR for proteins), one can detect the extent to which nucleotides in RNA are protected from cleavage reactions by hydroxyl radicals. It is known that the degree of protection is highly correlated with the solvent-accessible surface area (SASA) of the backbone sugar atoms, whereas there is no direct relationship between protection and its sequence and secondary structure \citep{Cate96S,Balasubramanian98P,Ding12NatMeth}. Consequently, the technique is useful for assessing the regions that are densely packed. In the context of ribozyme folding, it is often reported as ``protections'', which indicate regions where tertiary interactions form to a greater extent compared to some reference state, which is typically the unfolded state before folding is initiated. For the \Azo{} ribozyme, several footprinting data are available in the literature \citep{Adams04RNA, Rangan03PNAS, Rangan03JMB, Rangan04JMB, Chauhan08JACS, Chauhan09JMB, Duncan10PLoSOne, Mitra11RNA}. Because it is difficult to characterize the molecular details of the misfolded conformations in experiments, our simulations provide the needed quantitative insights into the protection factor at the nucleotide level, thus filling in details that cannot be resolved experimentally. 

We calculated the protection factors (footprints) for the native and the trapped ensembles based on SASA values of the simulated structures, and compared them with the experimental data. The two sets of footprints from the native and misfolded simulation ensemble have a similar pattern (Fig.~\ref{fig:Fig7}d) but with differing amplitudes, which is an indication of the extent of protection. Even though the two structural ensembles have different chain topologies, there are many well-packed regions in common. This also implies that the ensemble of misfolded conformations shares many common characteristics with the native structures, as implied by the kinetic partitioning mechanism (KPM) \citep{Guo95BP,Thirumalai96AccChemRes}. 

Experimental and calculated footprints are mostly consistent with the positions in both the protection factor profiles (Fig.~\ref{fig:Fig7}d). There are 28 nucleotides commonly protected in the three experimental data compared \citep{Adams04RNA, Rangan03PNAS, Chauhan08JACS}. Among these 28 nucleotides, 23 nucleotides are protected in the native ensemble (sensitivity (true positive rate) 0.82 and specificity (true negative rate) 0.79), whereas 24 nucleotides are protected in the misfolded ensemble from the simulation (sensitivity 0.86 and specificity 0.83), using a threshold protection factor, $F_{p}$ = 1.2.

The analysis based on Fig.~\ref{fig:Fig7}(d) quantitatively show that our simulation data and experiments are consistent with each other. However, the comparison also shows that footprinting analyses may not uniquely distinguish the native structure from the misfolded conformation unless the amplitudes are quantitatively compared.
Chauhan and Woodson \citep{Chauhan08JACS} (their footprinting data is shown in Fig.~\ref{fig:Fig7}d) reported that about 20\% of the population was misfolded, although these experiments cannot provide the molecular details of the misfolded structures. Given the good agreement found in Fig.~\ref{fig:Fig7}, we predict that the misfolded state identified in the simulations contributes to the 20\% fraction identified in experiments. The topologically-trapped states are more flexible than the native structure, which is reflected in the decreased amplitude in the protection factor (Fig.~\ref{fig:Fig7}d).

\section*{Discussion}
We performed coarse-grained simulations to reveal the structural details and the mechanisms by which specific and correlated association of \Mg{} with nucleotides drive the multistep folding kinetics of \Azo{} group-I intron RNA. The simulated collapse kinetics ($R_{g}$ versus time) and the tSAXS experimental data \citep{Roh10JACS} are in excellent agreement with each other. There are three major phases in the folding kinetics: (1) Rapid collapse from the unfolded ($\meanRg{} \approx7.8\,\textrm{nm}$) to an intermediate state in which the RNA is compact ($\meanRg{} \approx4\,\textrm{nm}$). (2) The second phase involves the formation of the \Ic{} state that is almost as compact as the native structure ($\meanRg{}\approx3.5\,\textrm{nm}$). (3) In the final phase, there is a transition to the native structure with $\meanRg{}\approx$~3~nm. Interestingly, most (about 80\%) of the secondary structures are formed within the first phase. In contrast, only about half of the tertiary interactions form incrementally during the first and second phases, and the remaining tertiary contacts form in the last phase. This suggests that the folding transition state is close to the folded state, which confirms the conjecture made previously \cite{Koculi06JMB}.

A surprising finding in our simulations is that \Mg{} ions condense onto the ribozyme over a very short time window, $t<$~0.05~ms, which results in the release of \K{}, an entropically favorable event. Nearly 90\% of \Mg{} ions are condensed in this time frame, even though most of the tertiary interactions and some helices are disordered. These findings show that \Mg{} condensation, in conjunction with \K{} release, precedes the formation of major ion-driven rearrangements in the ribozyme. We believe that this is what transpires in ribozymes and compactly folded RNAs.

\medskip{}

\textbf{Kinetic partitioning:}
The initial ribozyme collapse could be either specific, which would populate native-like structures that would reach the folded state rapidly, as predicted by the KPM \cite{Guo95BP,Thirumalai01AnnuRevPhysChem}, or it could be non-specific. In the latter case, the ribozyme would be kinetically trapped in the metastable structures for arbitrarily long times. In either case, theory has shown that the collapse time, $\tau_c \approx \tau_0 N^\alpha$ ($N$ is the number of nucleotides and $\alpha\approx1$) with the prefactor, $\tau_0$, that is on the order of $(0.1-1)\mu$s \cite{Thirumalai01AnnuRevPhysChem}. Taking $\tau_0\approx0.5\,\mu$s leads to the theoretical prediction that for the 195-nucleotide \Azo{} ribozyme, $\tau_c\approx(0.02-0.2)$ms, which is in accord with both simulations and experiments.

To determine if the structural variations at the earliest stage of folding affect the fate of the RNA, we analyzed the ensemble of conformations immediately after the initial collapse ($t < 150$ $\mu$s). Fig.~\ref{fig:Fig4}(c) shows the probability distributions of the structural overlap function ($\chi$) calculated for the {\it just-collapsed} ensemble. The order parameter, $\chi$, measures the extent of structural similarity to the native structure (0 for no similarity and 1 if it matched the folded state). The data is decomposed into four categories depending on the fate of each trajectory, rapidly folded, slowly folded, trapped, and trajectories that are neither folded nor kinetically trapped. There are two major findings in this plot: (1) Trajectories that result in either folded or trapped states have higher structural similarity to the folded state compared to those that do not reach these states at an early folding stage. (2) The distribution of the rapidly folded ensemble has a long tail with substantial similarity to the native structure, indicating that rapid folding to the folded structure arises from specific collapse. Although this result was expected on theoretical grounds \cite{Thirumalai95JdePhysique}, which has been established for RNA molecules whose folding rates vary over 7 orders of magnitude \citep{Hyeon12BJ}, there has been no direct demonstration of specific collapse and the associated structures until this study.

\medskip{}

\textbf{Mispaired secondary structures impede folding:}
We observed several incorrect base parings en route to either the native or the misfolded state (Figs.~S8 and S21). RNA sequences, in general, tend to form diverse secondary structures because there are likely multiple pairs of partially complementary regions. For instance, in mRNAs that do not have specific tertiary structures, many different patterns of secondary structures have been observed for a single sequence \citep{Garmann15RNA}. It is clear that even for a well-evolved sequence that has a specific tertiary structure, like the ribozyme, non-native complementary pairs can not be entirely avoided \citep{WuTinoco98PNAS}. Such mispaired helices inevitably slow down the folding \citep{Zarrinkar94Sci} and have to unfold, at least partially, before the RNA reaches the native state \citep{Wan10JMB}.

For group I intron ribozyme, it has been suggested that helix P3 has an alternative paring pattern (alt-P3) \citep{Pan98JMB}, which we observe as P3d-J8/7 in the simulations. The alt-P3 is thought to be a major reason for the slow folding rates of group I intron ribozyme. In accord with this proposal, it was shown that a point mutation in alt-P3, which stabilizes the correct pairing, increased the folding rate by 50 times in {\it Tetrahymena} ribozyme \citep{Pan00JMB}. In line with the same reasoning, \textit{Candida} ribozyme, which does not have stable alt-P3, folds rapidly without kinetic traps \citep{Zhang05RNA}. To our knowledge, other combinations of mispaired helices reported here (Fig.~S8) have not been detected in experiments, even though such mispaired helices are  permissible. There are a variety of metastable structures that render the folding landscape of RNA rugged, besides alt-P3. 

\medskip{}

\textbf{Topological frustration causes trapping in long-lived metastable states:}
We also found that the intermediate state, $I_{c}$, consists of not only on-pathway conformations but also misfolded structures.
The \Azo{} ribozyme folds correctly in a shorter time than topologically similar but larger ribozymes such as group I intron from \Tet{}. Nevertheless, several experiments have shown the existence of metastable states \citep{Chauhan09JMB,Roh10JACS,Sinan11JBC}. These metastable states slowly transition to the native state, which can be accelerated by urea \cite{Pan97JMB,Pan97NSB}. We did not see refolding events from the misfolded state to the native state within our simulation time (30 ms). This is also consistent with experiments, which showed that the misfolded state is long-lived and remained stable after 5 minutes of incubation with \Mg{} \citep{Sinan11JBC}. The estimated refolding rate was 0.29~min$^{-1}$ at 32$^{\circ}$C with 5~mM \Mg{}.

In our simulations, the ribozyme misfolded to a topologically frustrated state when one of the peripheral contacts, TL2-TR8 and TL9-TR5, formed early before the majority of the other tertiary contacts were fully formed (Fig.~\ref{fig:Fig4}). The reason for the more pronounced involvement of TL2-TR8 and TL9-TR5 in misfolding is that topological entanglement can easily occur when contacts in peripheral regions form first. We surmise that this mechanism is common to the folding of larger ribozymes.

Very few experiments have reported on the details of the misfolded structure due to the difficulty in distinguishing heterogeneous and transient structures. In \Tet{} ribozyme, it was suggested that the misfolded state has a similar topology to the native state, but with less non-native pairing \citep{Russell06JMB}. The misfolded state is mostly stabilized by native-like interactions, but there is ``strand-crossing'', by which the RNA conformation is trapped and unable to recover the native structure. Our results show that this is also the case in the smaller and faster-folding \Azo{} ribozyme. The metastable states found here consist of correct base pairs and tertiary interactions, which imply that they are native-like topological kinetic traps.

Recently, two groups independently reported cryo-EM structures of the topologically trapped state of the \Tet{} group I intron \cite{Bonilla22SciAdv, Li22PNAS}, that shares much in common with the architecture of \Azo{}. In both these structures, as predicted here, the J8/7 strand of the central core regions is in an incorrect position relative to the P3 and P7 helices. In Fig.~\ref{fig:Fig7}, a comparison of these structures with our simulated misfolded structure (J8/7) shows remarkable agreement in the strand arrangement. We surmise that our simulations, conducted without any prior knowledge of the topologically trapped states, predict the major structures of the metastable states that are populated during \Azo{} ribozyme folding. Our simulations  allow us to trace the process of this misfolding back to the intermediate state (Fig.~\ref{fig:Fig7}b), thus establishing a kinetic basis for their formation. Compared to the intermediate state in the correct folding pathway (Fig.~\ref{fig:Fig7}a), we now see how the subtle difference in the folding trajectory of the J8/7 strand ultimately leads to the native and topologically trapped native-like states, both of which are stabilized by the peripheral contacts in a similar way.

\medskip{}

\textbf{Role of monovalent ions:} \Azo{} ribozyme folds into a compact structure at high concentrations of monovalent ions even in the absence of \Mg{} \citep{Perez-Salas04Biochemistry}.  However, \Mg{} is essential for splicing activity \citep{Rangan03JMB}. Nevertheless, both previous experiments \citep{Perez-Salas04Biochemistry} and the recent empirical observation \citep{Rissone22PNAS, Bizarro12NAR} that roughly 1 mM \Mg{} is equivalent to 80 mM of monovalent cations raise the possibility that high monovalent cations could substitute for \Mg{}.  Therefore, it is interesting to pose the following question. Are the pathways explored by the ribozyme at high monovalent concentrations or when folding is driven by \Mg{} equivalent? This question can only be answered using simulations of the kind reported here. However, such simulations are demanding because large system sizes are needed to obtain reliable results. Despite the difficulties, this would be a problem worth investigating in the future.

\medskip{}

\textbf{Data Availability:}
The model structures are available in the online supplementary material. The simulation code and parameter files were deposited in Zenodo \url{https://doi.org/10.5281/zenodo.8246270}.

\medskip{}

\textbf{Supplementary Data:}
Supplementary Data are available at NAR online.

\medskip{}

\textbf{Funding:}
This work was supported in part by a grant from the National Science Foundation (CHE 2320256) and the Collie-Welch Regents Chair (F-0019) administered through the Welch Foundation.

\medskip{}

\begin{acknowledgments}
NH is grateful to Natalia Denesyuk for insightful discussions during the early stage of this study. We appreciate useful discussions with Sarah Woodson and Rick Russell. We thank Anne Bowen in the Texas Advanced Computing Center (TACC) at the University of Texas at Austin for rendering the simulation movies. We acknowledge the TACC for providing computational resources. 
\end{acknowledgments}

\bibliographystyle{nar}
\bibliography{AzoKinetics}

\end{document}


\title{Supplementary Information for ``Watching ion-driven kinetics of ribozyme folding and misfolding caused by energetic and topological frustration one molecule at a time''}

\author{Naoto Hori}
\email{hori.naoto@gmail.com}
\affiliation{Department of Chemistry, University of Texas, Austin, TX 78712, USA}
\affiliation{School of Pharmacy, University of Nottingham, Nottingham, UK}
\author{D. Thirumalai}
\email{dave.thirumalai@gmail.com}
\affiliation{Department of Chemistry, University of Texas, Austin, TX 78712, USA}
\affiliation{Department of Physics, University of Texas, Austin, TX 78712, USA}

\date{Aug 21, 2023}

\maketitle

\section{Supplementary methods}
\subsection{Coarse-grained RNA model with explicit ions}

\begin{figure}
\includegraphics[width=0.8 \textwidth]{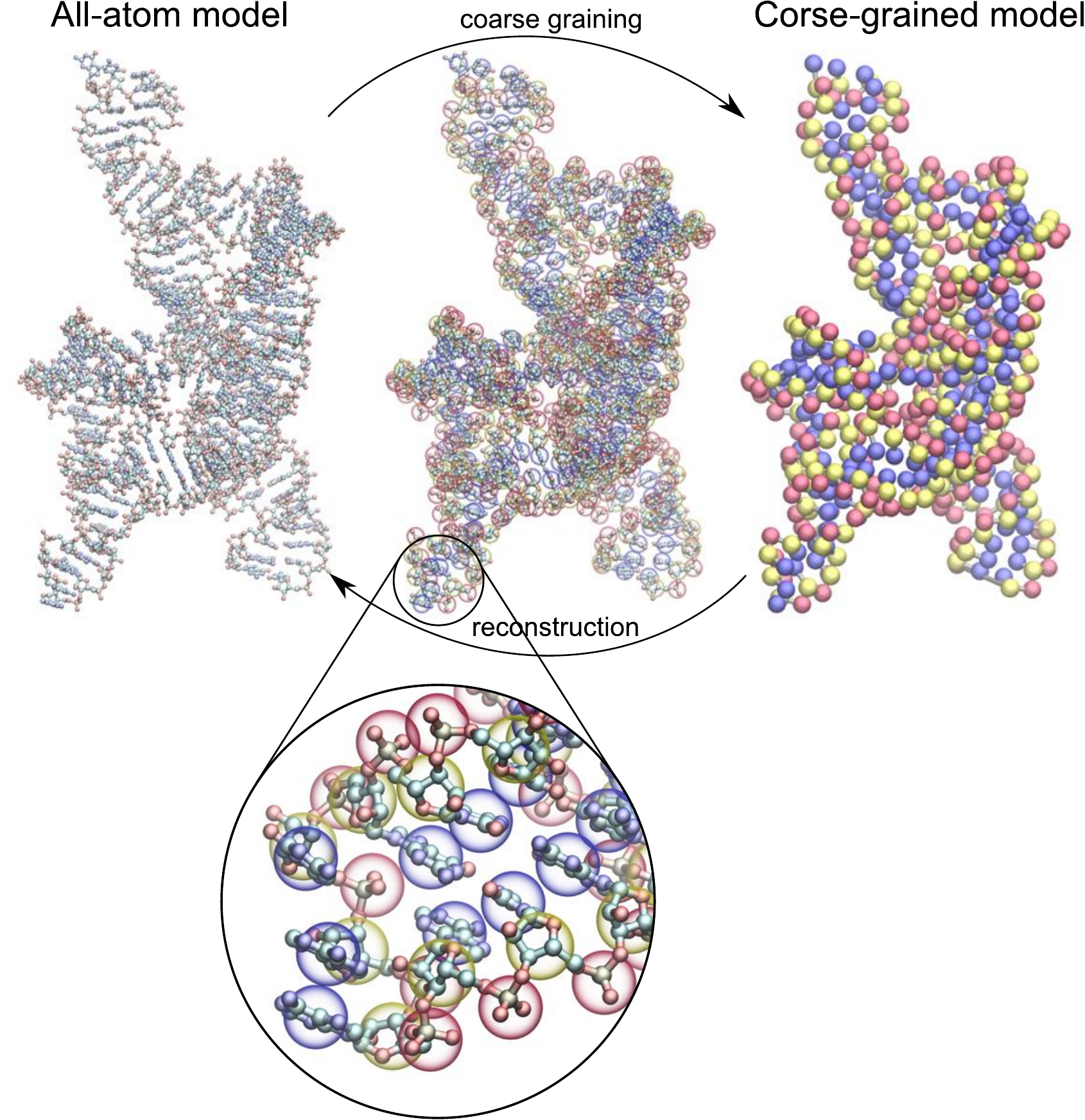}

\caption{
\label{fig:aa2cg}
\textbf{Three-Interaction-Site model.}
All-atom representation of \Azo{} group I intron RNA (left) is compared to the TIS coarse-grained representation (right). In the middle, the two representations are overlaid, and a part of the RNA is magnified at the bottom. The coarse-grained sites are colored in red (phosphate), yellow (sugar), and blue (base).
}
\end{figure}

We used the Three-Interaction-Site (TIS) RNA model \citep{Hyeon05P,Denesyuk15NatChem} in which each nucleotide is modeled using three spherical particles, corresponding to sugar (S), base (B), and phosphate (P) \citep{Hyeon05P} (Figure~\ref{fig:aa2cg}).
All the ions, \Mg{},  \K{}, and \Cl{}, including those in the buffer, are explicitly treated. The solvent (water) is treated implicitly using a temperature-dependent dielectric constant.  The TIS force field is written as, 
\begin{equation}
U_{\textrm{TIS}}=U_{\textrm{bond}}+U_{\textrm{angle}}+U_{\textrm{EV}}+U_{\textrm{HB}}+U_{\textrm{ST}}+U_{\textrm{ele}}.
\label{eq:U_TIS}
\end{equation}
The first two terms, $U_{\textrm{bond}}$ and $U_{\textrm{angle}}$ ensure the connectivity of the ribose backbone and bases with appropriate rigidity. They are modeled using harmonic potentials,
\begin{equation}
U_\textrm{bond}=\sum_{\textrm{bonds}}{k_b \left(\rho-\rho_0\right)^2}
\end{equation}
\begin{equation}
U_\textrm{angle}=\sum_{\textrm{angles}}{k_a \left(\theta-\theta_0\right)^2},
\end{equation}
where the reference values, $\rho_0$ and $\theta_0$, are taken from the ideal A-form RNA helix in a sequence-dependent manner.

\noindent {\it Excluded Volume Interactions:} The term $U_{\textrm{EV}}$ in equation (\ref{eq:U_TIS}) accounts for volume exclusion that prevents overlap between the CG sites. We used a modified version of the Lennard-Jones type potential,
\begin{eqnarray}
U_{\textrm{EV}}(r) & = & 
\begin{cases}
\sum_{(i+1<j)}{\varepsilon_{ij}\left[\left(\frac{a}{r+a-D_{ij}}\right)^{12}-2\left(\frac{a}{r+a-D_{ij}}\right)^{6}+1\right]} & (r\leq D_{ij})\\
0 & (r>D_{ij})
\end{cases}.
\end{eqnarray}
The exclusion distance is $D_{ij} = R_i + R_j$. The energy scale, $\varepsilon_{ij} = \sqrt{\varepsilon_i \varepsilon_j}$, is calculated using the standard combining rule. The values of $R_i$ and $\varepsilon_i$ are tabulated in Table S1. The choice of the distance $a=0.16$ nm ensures that $U_{\textrm{EV}}$ becomes a standard Lennard-Jones potential for a pair of interaction sites with the smallest particle type, \Mg{} ions.

\noindent {\it Hydrogen Bond Potential:} We consider hydrogen bonding interactions between bases for all possible canonical base pairs (any G-C, A-U, or G-U base pair can form during the folding process), as well as tertiary hydrogen bonds that are found in the crystal structure (PDB 1U6B \citep{Adams04RNA}). Tertiary hydrogen bonds can occur between any CG sites, whereas canonical base pair is formed only between two bases. Hydrogen bond potential is given by,
\begin{equation}
    U_{\textrm{HB}} = \sum_{\textrm{H-bonds}}{U_{\textrm{HB}}^{0}\exp{\left(-u_1\right)}} \label{eq:U_HB}
\end{equation}
where
\begin{eqnarray}
u_1 & = & \left(\begin{array}{c}
k_{r}(r-r_{0})^{2}+k_{\theta}(\theta_{1}-\theta_{1,0})^{2}+k_{\theta}(\theta_{2}-\theta_{2,0})^{2}\\
+k_{\psi}(\psi-\psi_{0})^{2}+k_{\psi1}(\psi_{1}-\psi_{1,0})^{2}+k_{\psi2}(\psi_{2}-\psi_{2,0})^{2}
\end{array}\right),
\label{eq:U_U1}
\end{eqnarray}
is a penalty for the deviation from the ideal hydrogen bonding configuration that is represented by the subscript of 0. The ideal configuration is taken from the A-form RNA double helix for the canonical base pairs, and from the crystal structure of the ribozyme (PDB 1U6B \citep{AzoXray04N}) for tertiary hydrogen bonds. The list of tertiary hydrogen bondings was generated by the WHAT-IF server \citep{WHATIF}.

\noindent {\it Base Stacking Potentials:} Stacking interactions consist of two types, $U_{\textrm{ST}} = U_{\textrm{SST}} + U_{\textrm{TST}}$. The first type, $U_{\textrm{SST}}$, is the sum of secondary base stacking that are interactions between all consecutive nucleotides, $i$ and $i+1$, regardless of whether they are involved in stacking interactions in the crystal structure. Tertiary stacking interaction, $U_{\textrm{TST}}$, is the sum of the base-stacking interactions between non-consecutive nucleotides, $i$ and $j$ ($|i-j|>1$), found in the crystal structure. The list of tertiary base-stacking interactions for {\Azo} ribozyme is given in Table S1 of Ref. \citep{Denesyuk15NatChem}.
The total stacking-interaction potential is given by,
\begin{equation}
    U_{\textrm{ST}}=\sum_{\textrm{secondary}}{U^{0}_{\textrm{SST}}\left(1+u_2\right)^{-1}} + \sum_{\textrm{tertiary}}{U^{0}_{\textrm{TST}}\left(1+u_1\right)^{-1}},
\end{equation}
where $U^0_{SST}$ and $U^0_{TST}$ are stability parameters for secondary- and tertiary-stacking interactions, respectively. For secondary stacking, the penalty term is, 
\begin{equation}
    u_2 = k_{r}(r-r_{0})^{2}+k_{\phi}(\phi_{1}-\phi_{1,0})^{2}+k_{\phi}(\phi_{2}-\phi_{2,0})^{2}.
\end{equation}
The value of $u_1$ for tertiary stacking is the same form as in equation (\ref{eq:U_U1}) for the hydrogen-bonding interactions.

\noindent {\it Electrostatic Potential:} The last term in equation (\ref{eq:U_TIS}) is the Coulomb potential for the electrostatic interactions,
\begin{equation}
U_{\textrm{ele}}=\frac{e^{2}}{4\pi\varepsilon_{0}\varepsilon}\sum_{i<j}\frac{Q_{i}Q_{j}}{r_{ij}}
\end{equation}
where $e$ is the elementary charge, $\varepsilon_{0}$ is the vacuum permittivity, and $Q$ is the charge of each ion type. We used the experimentally-determined temperature-dependent dielectric constant \citep{Malmberg1956},
\begin{equation}
\varepsilon(T_{c})=87.74-0.4008T_{c}+9.398\times10^{-4}T_{c}^{2}-1.41\times10^{-6}T_{c}^{3},
\end{equation}
where $T_{c}$ is the simulation temperature in degrees Celsius.
The ions interact with each other and with the RNA via the excluded volume and electrostatic interactions ($U_{\textrm{EV}}$ and $U_{\textrm{ele}}$), whereas the other four terms account for interactions within the RNA molecule.

In the previous study \citep{Denesyuk13JPCB, Denesyuk15NatChem}, the parameters in the TIS force field were optimized so that the model reproduced the thermodynamics of dinucleotides, several types of hairpins, and an RNA pseudoknot. The model with the optimized parameters also quantitatively reproduced the measured equilibrium properties  of \Azo{} ribozyme, such as the equilibrium $R_\textrm{g}$ as a function of \Mg{} concentrations, and \Mg{} binding sites \citep{Denesyuk15NatChem,Hori19BJ}. In addition, we obtained excellent agreements with experiments for heat capacities of pseudoknots and hairpins \citep{Denesyuk13JPCB, Denesyuk15NatChem}. These successful applications for calculating the equilibrium properties of RNA molecules set the stage for probing the more difficult problem of calculating the folding kinetics of a large ribozyme.

Note that we used a variant of the TIS model \citep{Denesyuk15NatChem}, not the one published later \citep{Nguyen19PNAS}. In the 2019 model \citep{Nguyen19PNAS}, we developed a theory to treat monovalent ions implicitly whereas the divalent ions were treated explicitly. Although the latter model is advantageous in terms of computational cost due to the absence of explicit monovalent ions, it was essential for us to use the explicit representations of both divalent and monovalent ions in the current study. This allowed us to directly quantify the condensation and binding of both monovalent and divalent ions (Figure 6a), and to vividly capture the replacement of \K{} ions by \Mg{}. This would not be possible using the implicit monovalent ion model.

\subsection{Simulation protocols}

In order to generate the folding trajectories, we performed Brownian dynamics simulations. The equation of motion is \citep{Ermak78JCP},
\begin{equation}
\dot{\boldsymbol{x}}=-\frac{1}{\gamma}\frac{\partial U_{\textrm{TIS}}}{\partial\boldsymbol{x}}+\boldsymbol{\Gamma}.\label{eq:Brownian}
\end{equation}
where $\boldsymbol{x}$ is a coordinate, and $\boldsymbol{\Gamma}$ is a Gaussian random force that satisfies the fluctuation-dissipation relation given by $\left\langle \boldsymbol{\Gamma}_{i}(t)\boldsymbol{\Gamma}_{j}(t')\right\rangle =6\gamma k_{\mathrm{B}}T\delta(t-t')\delta_{ij}$.
The friction coefficient follows the Stokes-Einstein relation, $\gamma=6\pi\eta R$, where $\eta = 8.9\times10^{-4}\,\textrm{Pa\ensuremath{\cdot}s}$ is the viscosity of water, and $R$ is appropriate radius for each coarse-grained site. We set values of $R$ for the RNA sites and ions according to previous studies~\cite{Volkov97, Fernandes02NAR}. 

The simulation time was mapped to the real time by relating the microscopic time scale $\tau_{L} = \sqrt{ma^2/\epsilon} \sim$ 2 ps and the characteristic time of the Brownian dynamics, $\tau_{H}=\frac{6\pi\eta a^3}{k_{B}T}\approx300$~ps, where the typical energy scale $\epsilon \sim1$ kcal/mol and the length scale $a\sim0.4$ nm \cite{Veitshans97,Hyeon08JACS}. We previously demonstrated that this way of predicting kinetics using the TIS model provides {\it quantitative} agreement for the time scale of RNA folding for hairpin and pseudoknot RNA molecules \citep{Cho09PNAS,Biyun11JACS,Roca18PNAS,Hori18JPCB}.

For each initial conformation, taken from the unfolded ensemble generated as described above, \Mg{} and \Cl{} ions corresponding to 5 mM MgCl$_2$ were added before initiating the folding simulations. The additional ions were randomly distributed in the cubic box, keeping them at least 0.5 nm from the preexisting ions and the ribozyme. 
The folding simulations were continued until the RNA reached the native conformation, or the time steps reached $10^{10}$ corresponding to $30\,\textrm{ms}$, which is the maximum simulation time in all our trajectories. If {\it Azoarcus} ribozyme did not fold within the maximum simulation time, then it is considered to be kinetically trapped, a phenomenon that frequently occurs in {\it in vitro} experiments.

The coordinates of the RNA and ions were recorded every 50,000 time steps. We assessed if RNA is folded within the time scale of simulations using the following criterion: Root mean square deviation (RMSD) from the PDB structure was computed every frame. The RMSD is then averaged over a time window of 1,000 frames. If the window average of RMSD is smaller than 0.6 nm, the trajectory is deemed to reach the folded state.
In a few trajectories, we found unphysical chain crossing when the length of the pseudo-covalent bond between sugar and phosphate becomes comparable to the excluded distance between them. Although this rarely occurred, it should not be dismissed because such crossing could affect the folding kinetics. Thus, when a chain crossing occurred in a trajectory, we restarted the simulation with a different random number seed to ensure the backbone of RNA does not cross each other.

\subsection{Analyses}
\textbf{Finding major misfolded conformations:} The clustering analysis was performed as described in Models and Methods in the main text.
The result of the clustering analysis is shown in Figure~\ref{fig:dendrogram} as a dendrogram along with structures of cluster centers. The compact-structural ensemble is grouped into five parent clusters that are designated as cluster identification (ID) 1 through 5 in the dendrogram. From a visual inspection of the structures, we identified the five clusters as follows: Cluster 1 is Misfolded (P2) conformation (see the main text), Cluster 2 belongs to the native-like ensemble, Clusters 3 and 5 are both Misfolded (J8/7) conformation, and Cluster 4 is another minor misfolded state that appeared in only one trajectory. Although structures in Clusters 3 and 5 are classified into two distinct clusters, they have the same chain topology and tertiary contacts, except that TL9-TR5 contact is formed in Cluster 3, whereas it is disrupted in Cluster 5. For sub-clusters 1a and 1b, 2a and 2b, and so on, we confirmed that they represent minor variations of the root clusters. This method provides an unbiased way to classify the misfolded structures, which are, except in rare cases (Cluster 1), difficult to observe in experiments. 

\medskip{}



\textbf{Calculation of \Mg{} binding rate:}
The mean first passage time (MFPT) of \Mg{} to bind to each phosphate site was calculated as follows. The phosphate associated with nucleotide site $i$ in a trajectory $j$ is considered to be bound by \Mg{} if the distance to the closest \Mg{} ion is shorter than the cutoff distance in 10 consecutive time frames ($1.5~\mu s$). The time at which the binding occurs for the first time, $t_{i,j}$, was recorded as the first passage time of the trajectory. The MFPT for the $i$\textsuperscript{th} nucleotide was calculated as,
\begin{equation}
    \tau_{i} = \frac{1}{N}\sum_j^{N}{t_{i,j}},\label{eq:FPT}
\end{equation}
where $N$ is the number of trajectories.
The \Mg{} binding rate of the $i$\textsuperscript{th} nucleotide is,
\begin{equation}
    k_{b}(i) = \frac{1}{\tau_{i}}.
\end{equation}
Note that, for some nucleotide sites in certain trajectories, \Mg{} ions do not bind tightly within the simulation time $T_{s}$. In such cases, we assign $t_{i,j} = T_{s}$ in equation (\ref{eq:FPT}). In that sense, $k_{b}$ is an estimate of the upper bound of the binding rate from the simulations because the true first passage time ($t_{i,j}$) would be larger in such trajectories.

\medskip{}

\textbf{Definition of the folded helix, tertiary interactions, and non-native helix:} The fraction of helix formation (Figure~\ref{fig:Fig5}e) was calculated as the fraction of base pairs formed in the helix at a specific time. A base pair was assumed to be formed if the magnitude of the hydrogen-bonding energy (equation~\ref{eq:U_HB}) was more favorable than the thermal energy, $k_{\textrm{B}}T$. 
Similarly, the formation of key tertiary interactions (Figure~\ref{fig:Fig1} and \ref{fig:Fig5}f) was assessed by comparing the tertiary hydrogen-bonding or stacking energy to the thermal energy.

In order to find the major mispaired helices, \textit{i.e.} double-stranded helices consisting of non-native base pairs shown in Figure~\ref{fig:mispairing}, we first made a list of non-native base pairs, that contained all possible canonical base pairs (A-U, G-C, and G-U) that are not present in the crystal structure. Then we counted the frequencies of each non-native base pair over all the trajectories. Lastly, we sorted the list by the frequencies of their formation. The major mispaired helices were identified from combinations of base pairs that formed most frequently, which appeared at the top of the sorted list. The fractions of mispaired helices (Figure~\ref{fig:nnpair_fraction}) were calculated by checking whether the mispaired helices are formed or not at each time step in each trajectory. If two or more non-native base pairs formed, the helix was deemed to be formed.

\section{Supplementary discussion}

\textbf{Folding kinetics of helices:}
To investigate the details of the folding process, we focussed on the formation of individual secondary structural elements. We calculated the time-dependent folded fraction of helices using the trajectories which reached the native structure (Figure~\ref{fig:Fig5}e). There are eight helices, P2 through P9 in the \Azo{} ribozyme (Figure~\ref{fig:Fig1}a). At $t=0$ and in the absence of \Mg{}, helices P2, P4, P5, and P8 are stably folded, which are also found in equilibrium simulations (Figure~\ref{fig:Fig5}e). The presence of stable helix regions, including the partial formation of P6 and P9 (see the following), is consistent with experimental results \citep{Rangan03PNAS}. Consequently, the fractions of these four helices start from high (> 80\%) at $t=0$, and gradually increase throughout the folding process. A part of the P9 helix can be sectionalized as P9.0, which is not stable in the absence of \Mg{}.  Due to the unstable part, the fraction of P9 is about $\sim$0.7 in the early stage of folding (purple line in Figure~\ref{fig:Fig5}).  Nevertheless, the entire P9, including P9.0, formed rapidly in $t<0.1\,\textrm{ms}$ upon titrating with 5 mM \Mg{}. Helix P6 also has a sub-region, P6a, which is more stable than the rest. The six base-pairs at the tip of P6 are stable even in the absence of \Mg{}, contributing to about 40\% formation of the helix in the initial stage (light green in Figure~\ref{fig:Fig5}). The other base pairs in P6 fold on a longer time scale. Formation of the three base pairs involved in the Tertiary-Helix (TH, see \ref{fig:Fig1}b) is particularly slow; it folds in the final stages, $t>10\,\textrm{ms}$. 

Even though some parts of P6 and P9 fold more slowly than others, helices P2, P4, P5, P6, P8, and P9 are essentially simple stem-loop structures, with the two strands being located close to each other along the sequence. Therefore, we surmise that there is a negligible entropic barrier for their formation so that they fold rapidly (or even form, to a great extent, in the absence of \Mg{}). Helices P3 and P7 exhibit particularly slow folding kinetics (light and dark blues in Figure~\ref{fig:Fig5}). The two strands of P3 and P7 helices are far along the sequence, and thus it takes more time to search for their counterparts.

\medskip{}

\textbf{A minor topological trap by incorrect positioning of the P2 helix:}
In addition to the major topologically trapped state (Misfold J8/7), described  in the main text, we found another minor topological misfolding due to incorrect positioning of the P2 helix domain. In this Misfold P2, one of the P3 strands leading to the P2 domain is positioned in the opposite direction relative to its native position when the TL9-TR5 tertiary interaction forms, thus fixing the P2 domain on the wrong side (Figure~\ref{fig:P2misfold_intermediate}). The space between the two pairs of coaxial helix domains P4-P5-P6 and P7-P8-P9 is too narrow for the P2 helix to move to its normal position on the opposite side unless TL9-TR5 dissociates.

By searching through all the folding trajectories using the representative structure from the clustering analysis, we found that the misfolded state occurred in only three out of 95 trajectories. In two of the three trajectories, the TL9-TR5 interaction persisted after the formation of the P2 misfolded state, and therefore the ribozyme remained trapped throughout the simulation (Figure~\ref{fig:P2misfold_trapped}). In the other trajectory, the TL9-TR5 interaction was eventually disrupted by thermal fluctuations and the P2 misfolded state was resolved (Figure~\ref{fig:P2misfold_resolved}). However, the ribozyme did not fold to the native structure within the simulation time. Although it is difficult to deduce the exact mechanism by which this misfolded state occurs from only three trajectories, what is common in these trajectories is the partial unfolding of the P6 helix and the relatively early formation of the G site (Figs.~\ref{fig:P2misfold_trapped} and \ref{fig:P2misfold_resolved}).

\medskip{}

\textbf{TH formation and \Mg{} binding to U51 and U126:}
Triple Helix (TH) is a structural element that folds in the early stage following Stack Exchange (Figure~\ref{fig:Fig5}f). As shown in Figure~\ref{fig:cor_TH}, nucleotides U51 and U126 bind \Mg{} ions upon folding of the TH, which is consistent with the observation that they are coordinated by two \Mg{} ions in the crystal structure. The correlation coefficients of the first passage times are relatively low, $\rho = 0.58$ and $0.73$ for U51 and U126, respectively. The scatter plots in Figure~\ref{fig:cor_TH}(c, d) show that, in some trajectories, TH forms earlier than \Mg{} binding to U51 and U126, indicating that \Mg{} binding to these nucleotides are not a strict requirement for TH to form. This is consistent with the finding that TH forms at high probability ($\approx$0.6) even at submillimolar \Mg{} concentration at equilibrium \citep{Denesyuk15NatChem}. 

\textbf{Correlation between SE formation and \Mg{} binding to G41, G138 and A167-A168:}
The Stack Exchange junction (SE) is the coaxial stacking between P3 and P8 helices, whose formation facilitates the folding of the P3 pseudoknot. Figure~\ref{fig:cor_SE} shows that the formation of SE temporally correlates with the \Mg{} binding at a cavity formed by G41, G138, and A167-A168. The correlation coefficients are all above 0.9. Based on the result of equilibrium simulations \citep{Denesyuk15NatChem}, it was conjectured that \Mg{} ions bind this region in a diffusive manner, which makes it difficult to resolve \Mg{} densities in the crystal structures. The results of our kinetics simulations also support the finding that \Mg{} binding events to these nucleotides are important for the formation of the SE motif.

\medskip{}

\textbf{Correlation between \Mg{} binding and formations of peripheral interactions:}
Two peripheral tertiary motifs, TL2-TR8 and TL9-TR5, are also stabilized by \Mg{} ions. As shown in Figure~\ref{fig:cor_TL2TR8} and \ref{fig:cor_TL9TR5}, both peripheral motifs have two or three nucleotides having high temporal correlations between \Mg{} binding and formations of the motif. There is, however, a qualitative difference between the two cases. In the TL2-TR8 case, the tertiary interaction may form earlier than the time for specific \Mg{} binding events, although they are at the same time in the majority of the trajectories (Figure~\ref{fig:cor_TL2TR8} a-b). We did not observe any trajectory in which \Mg{} bound before the motif was formed. In contrast, \Mg{} binding to TL9-TR5 motif can be earlier or later than the formation of the interaction (Figure~\ref{fig:cor_TL9TR5} a-c). These events are temporally correlated. The Pearson correlation coefficients of the first passage times are higher for TL9-TR5 ($\rho = 0.79$, $0.82$, $0.97$) than for TL2-TR8 ($\rho = 0.58$ and $0.73$). This may be related to the subtle difference in their stability observed in equilibrium simulations, that is, TL9-TR5 unfolded rapidly at low \Mg{} concentrations (1 mM) whereas TL2-TR8 was still partially folded ($\sim30$\% population) \citep{Denesyuk15NatChem}.

\clearpage{}

\bibliographystyle{nar}
\bibliography{AzoKinetics}

\clearpage{}

\begin{table}
\caption{
\label{tab:radii}
\textbf{Parameters for excluded volume and electrostatic interactions.}
In the first column, P is phosphate, S is sugar, and A, G, C, and U are the four bases.
The charge, $Q_{i}$, is defined in units of the proton charge.
If both interacting sites are RNA sites, we take $R_{i} + R_{j} = 0.32$ nm.
}
\medskip{}
\setlength{\tabcolsep}{1em}
\begin{tabular}{  c | c | c | c  } 
 \hline\hline
 & $R_{i}$, nm & $\varepsilon$, kcal/mol & $Q_{i}$ \\
 \hline
 P & 0.21 & 0.2 & -1 \\ 
 S & 0.29 & 0.2 & 0 \\  
 A & 0.28 & 0.2 & 0 \\  
 G & 0.30 & 0.2 & 0 \\  
 C & 0.27 & 0.2 & 0 \\  
 U & 0.27 & 0.2 & 0 \\  
 \Mg & 0.08 & 0.9 & 2 \\  
 \Cl & 0.19 & 0.3 & -1 \\ 
 \K & 0.27 & 0.0003 & 1 \\ 
 \hline\hline
\end{tabular}
\end{table}

\begin{figure}
\includegraphics[width=0.5\textwidth]{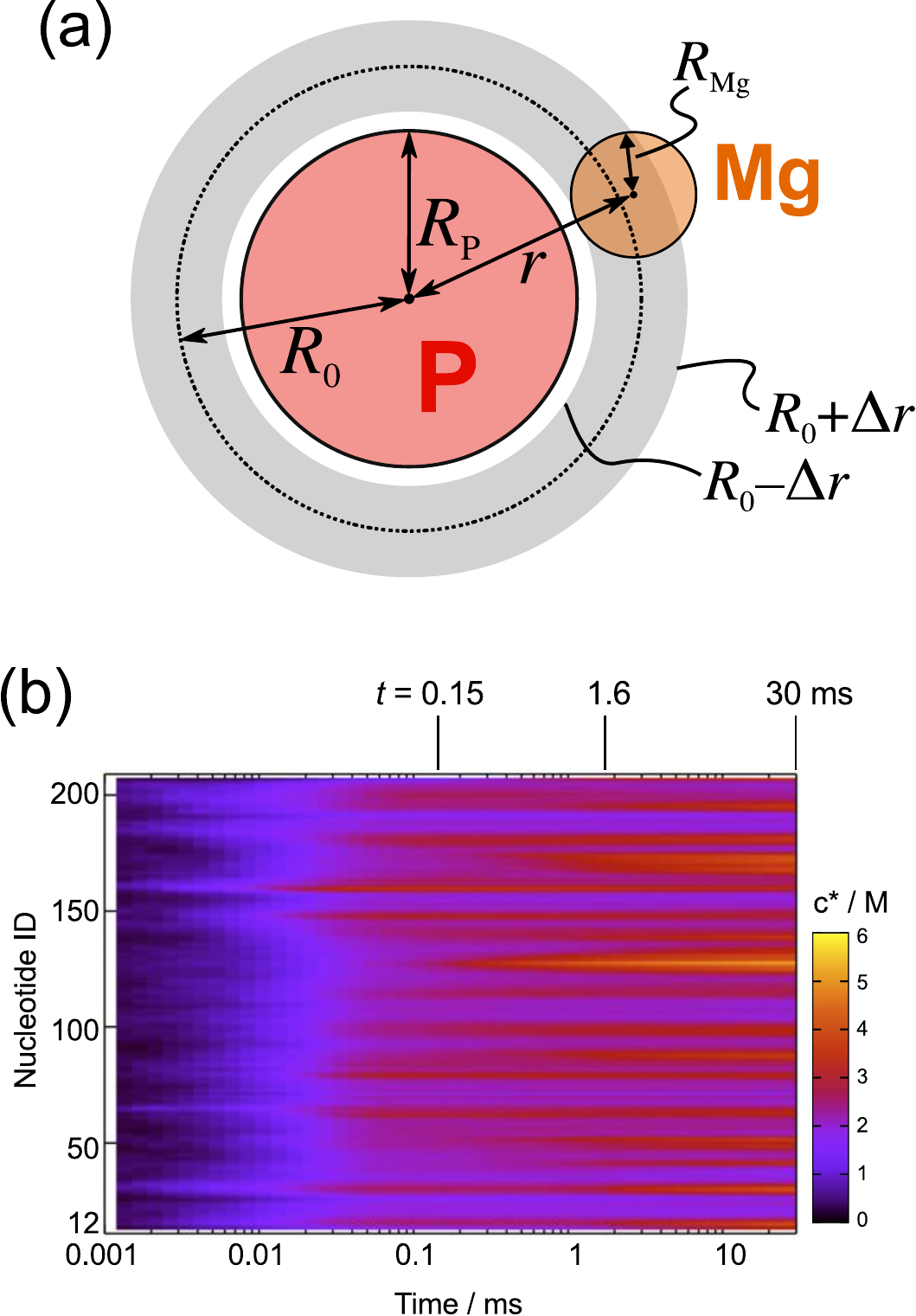}

\caption{
\label{fig:Mgbinding_def_and_3D}
\textbf{Time dependent binding of \Mg{}.}
\textbf{(a)} A schematic cross-section of a \Mg{} ion bound to a phosphate group (P), used to define the contact \Mg{} concentration, $c^{\ast}$.
We counted the number of \Mg{} ions located in the range, $r_0 - \Delta r < r < r_0 + \Delta r$, where $r$ is the distance from the phosphate, $r_0 = R_{\textrm{P}} + R_{\textrm{Mg}}$ is the sum of the excluded-volume radii, and $\Delta r = 0.15$ nm is a margin. The contact \Mg{} concentration is then calculated by dividing the number by the spherical shell volume (shaded in grey).
\textbf{(b)} Average contact \Mg{} concentration at each nucleotide as a function of time. In the main text, \Mg{} fingerprints are shown as the profile of $c^{\ast}$ at $t=0.15$, 1.6, 30 ms (Figure~\ref{fig:Fig6}). The heterogeneity of association is considerable.}
\end{figure}

\begin{figure}
\includegraphics[height=0.7\textheight]{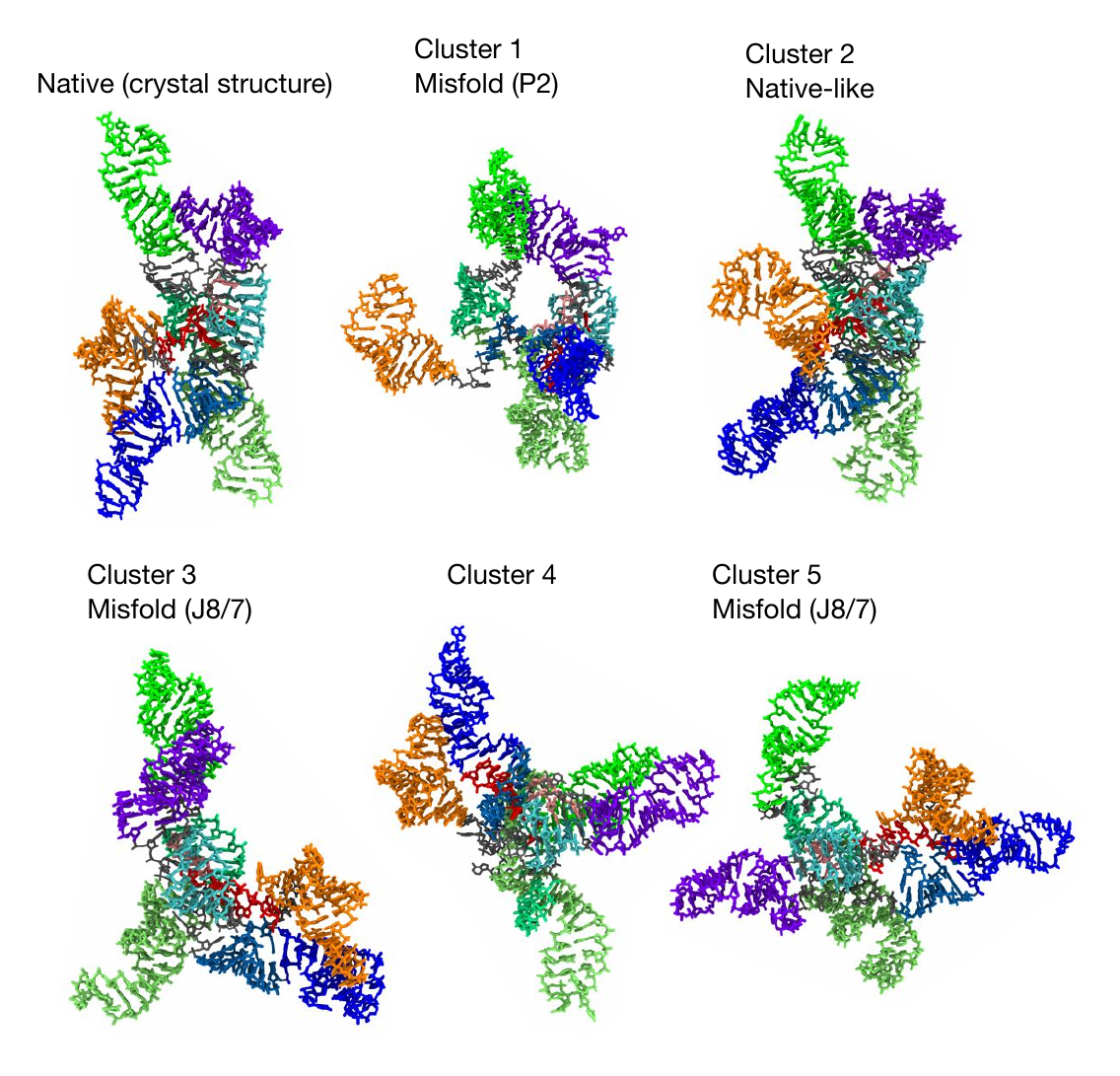}

\caption{
\label{fig:misfolded_aa_structures}
\textbf{Atomistic representations of cluster-center structures.}
The same representative misfolded structures as shown in Figure \ref{fig:dendrogram} with the native structure (top left). The structures are reconstructed from coarse-grained coordinates using the fragment-assembly approach followed by energy minimization, as described in Methods. 
}
\end{figure}

\begin{figure}
\includegraphics[width=0.6\textwidth]{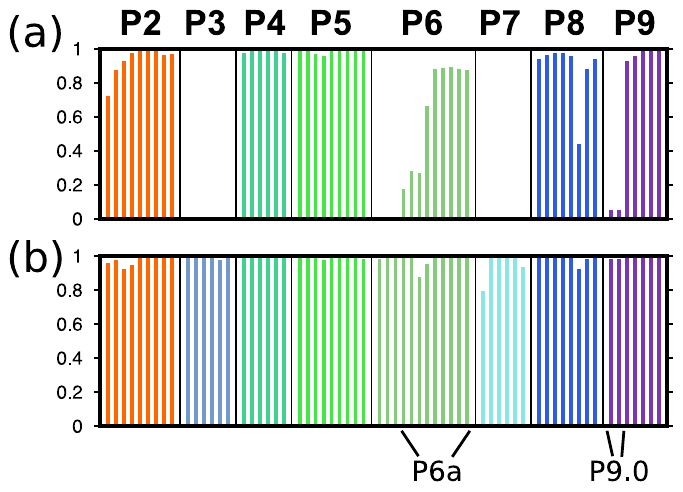}

\caption{
\label{fig:helix_equil}
\textbf{}
Probabilities of helix formation for helices P2 to P9 under equilibrium conditions in the (a) absence and (b) presence (5~mM) of \Mg.
The labels for helices are shown on top of each block.
Each bar represents a base pair consisting of each helix. P6 and P9 are sectioned into parts, P6a and P9.0, which are indicated in the bottom panel (see also the secondary structure in Figure 1a in the main text).
}
\end{figure}

\begin{figure}
\includegraphics[width=0.9\textwidth]{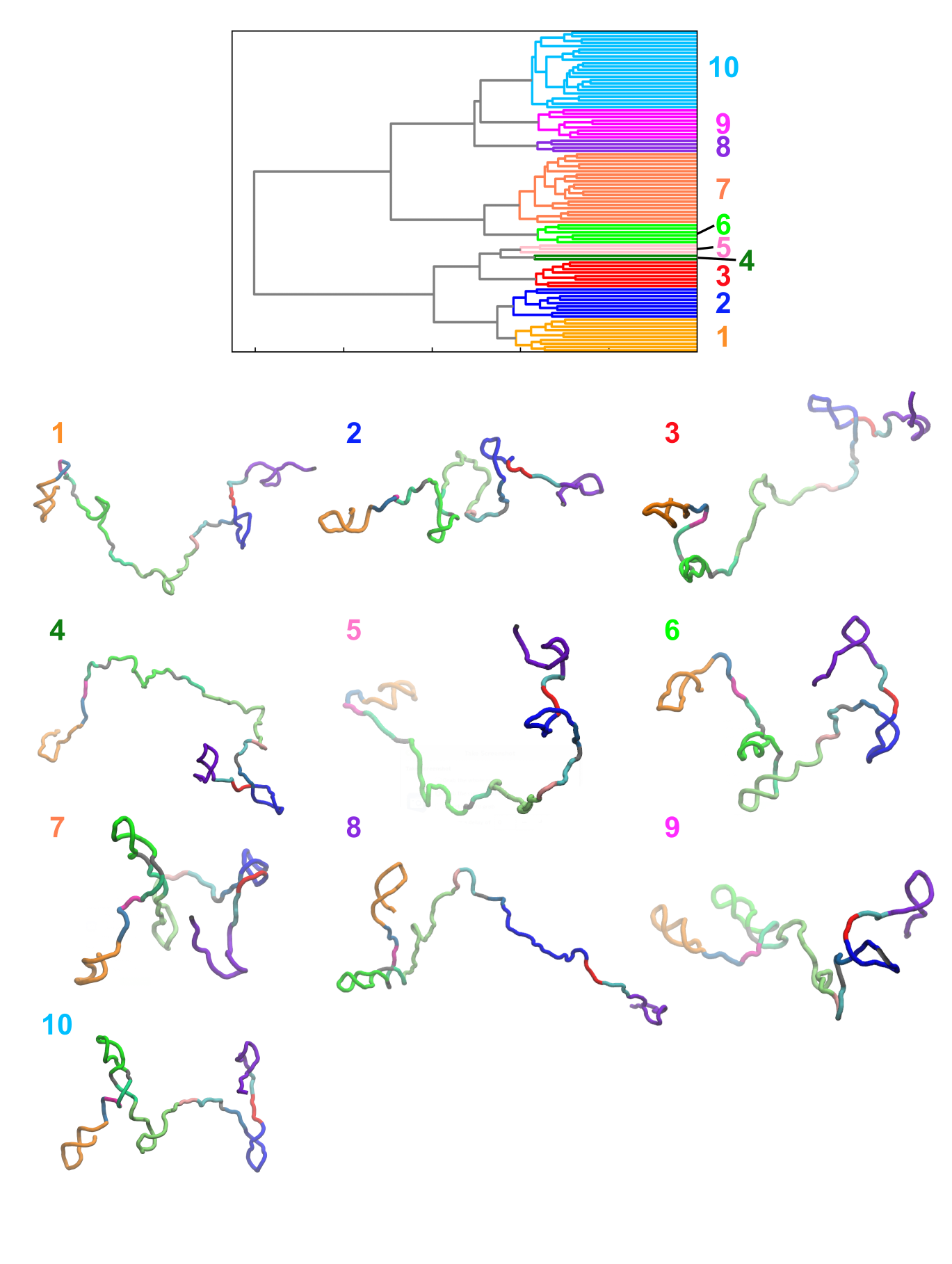}

\caption{
\label{fig:cls_initials}
\textbf{Hierarchical clustering of initial structures.}
The dendrogram (top) shows the result of clustering analysis for the 95 initial structures with DRID as the distance metric, in the same manner as the clustering of compact structures. Representative structures of the top 10 clusters are shown. The unfolded conformations are structurally diverse.
}
\end{figure}

\begin{figure}
\includegraphics[]{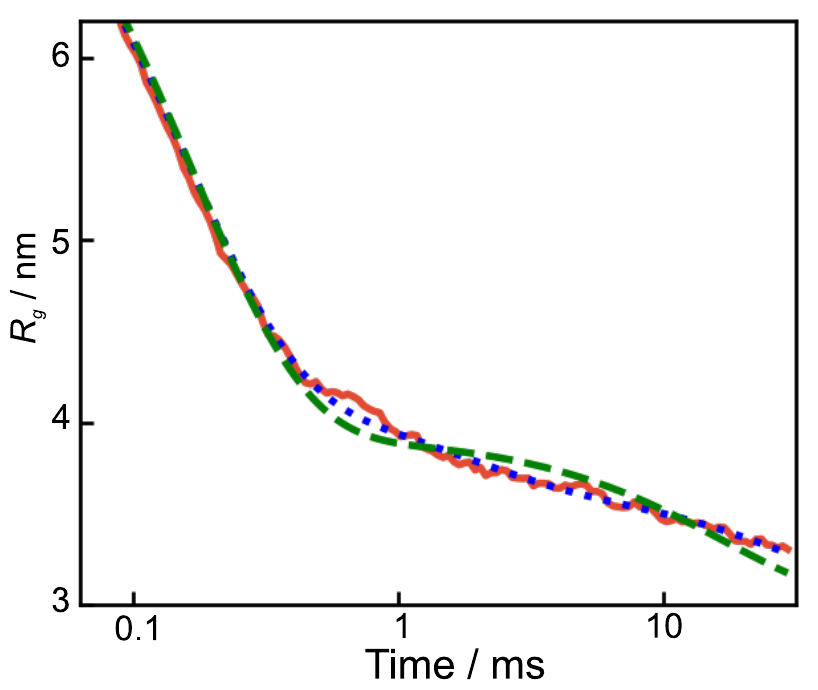}

\caption{
\label{fig:Rg_fit}
\textbf{Fits of ensemble-averaged radius of gyration ($R_\textrm{g})$ as a function of time.}
The solid red line is $\left<\Rg{}\right>$, calculated from all the simulation trajectories. The green dashed line is the best fit using a sum of two exponential functions ($\Phi_1 = 0.81$, $\Phi_2 = 0.19$, with the corresponding time constants, $\tau_1 = 0.17$, $\tau_2 = 17$ ms). The blue dotted line is the fit using three exponential functions discussed in the main text. The sum of two exponential functions does not quantitatively capture the time dependence beyond $\sim$1 ms. There is a noticeable deviation between the red and green curves.
}
\end{figure}

\begin{figure}
\includegraphics[height=0.7\textheight]{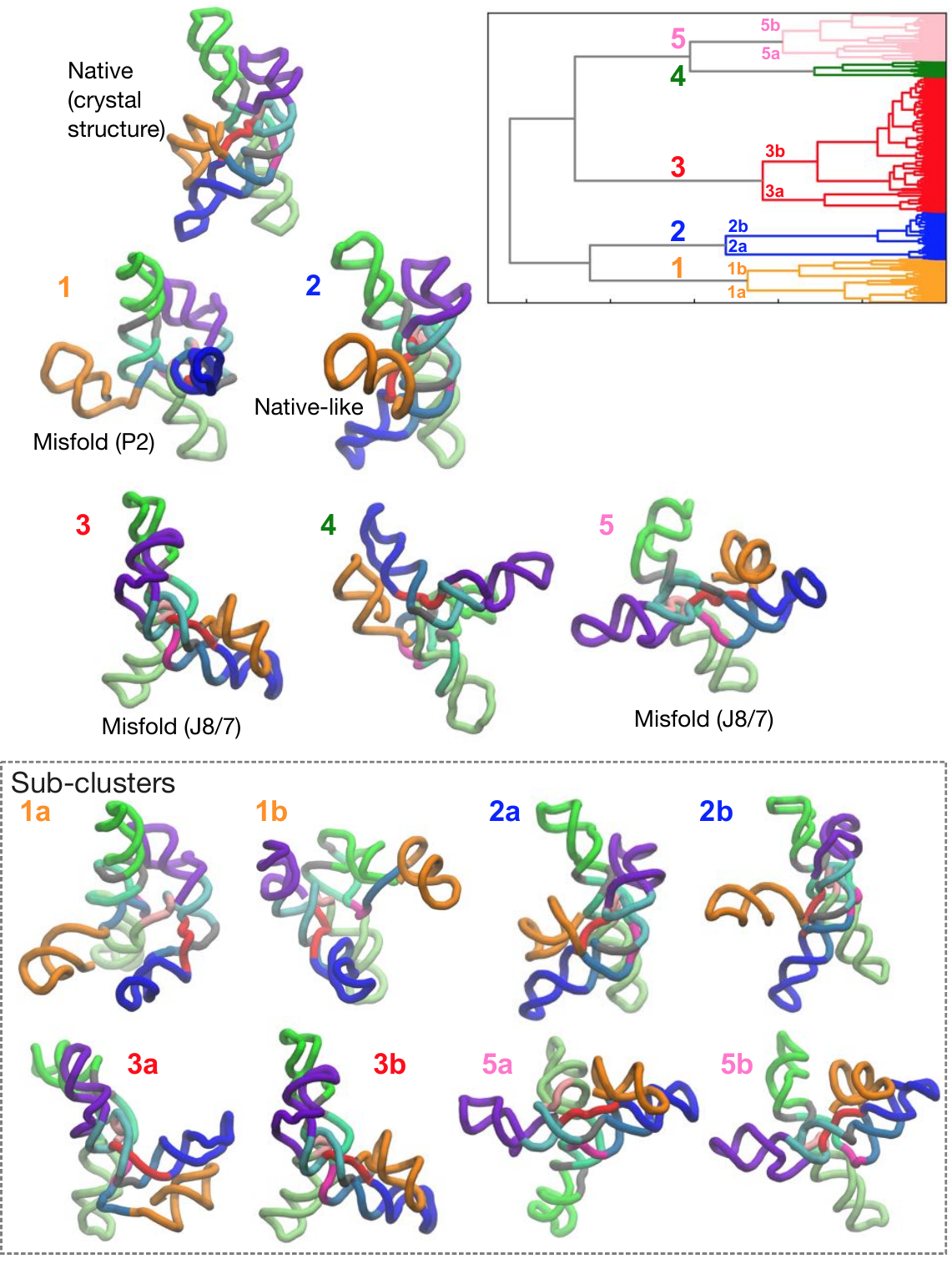}

\caption{
\label{fig:dendrogram}
\textbf{Clustering of compact structures from the simulation trajectories.}
(Top right) The dendrogram from the hierarchical clustering based on pairwise DRID. Cluster IDs are assigned for several representative clusters.
(Top left) The crystal structure of the ribozyme is shown for comparison.
(Middle) Structures from the five major cluster centers.
Cluster 2 contains the native topology conformations.
The other four clusters are misfolded conformations.
(Bottom) Structures of sub-clusters of Cluster 1, 2, 3, and 5 are also shown for comparison.
}
\end{figure}

\begin{figure}
\includegraphics[width=0.9\textwidth]{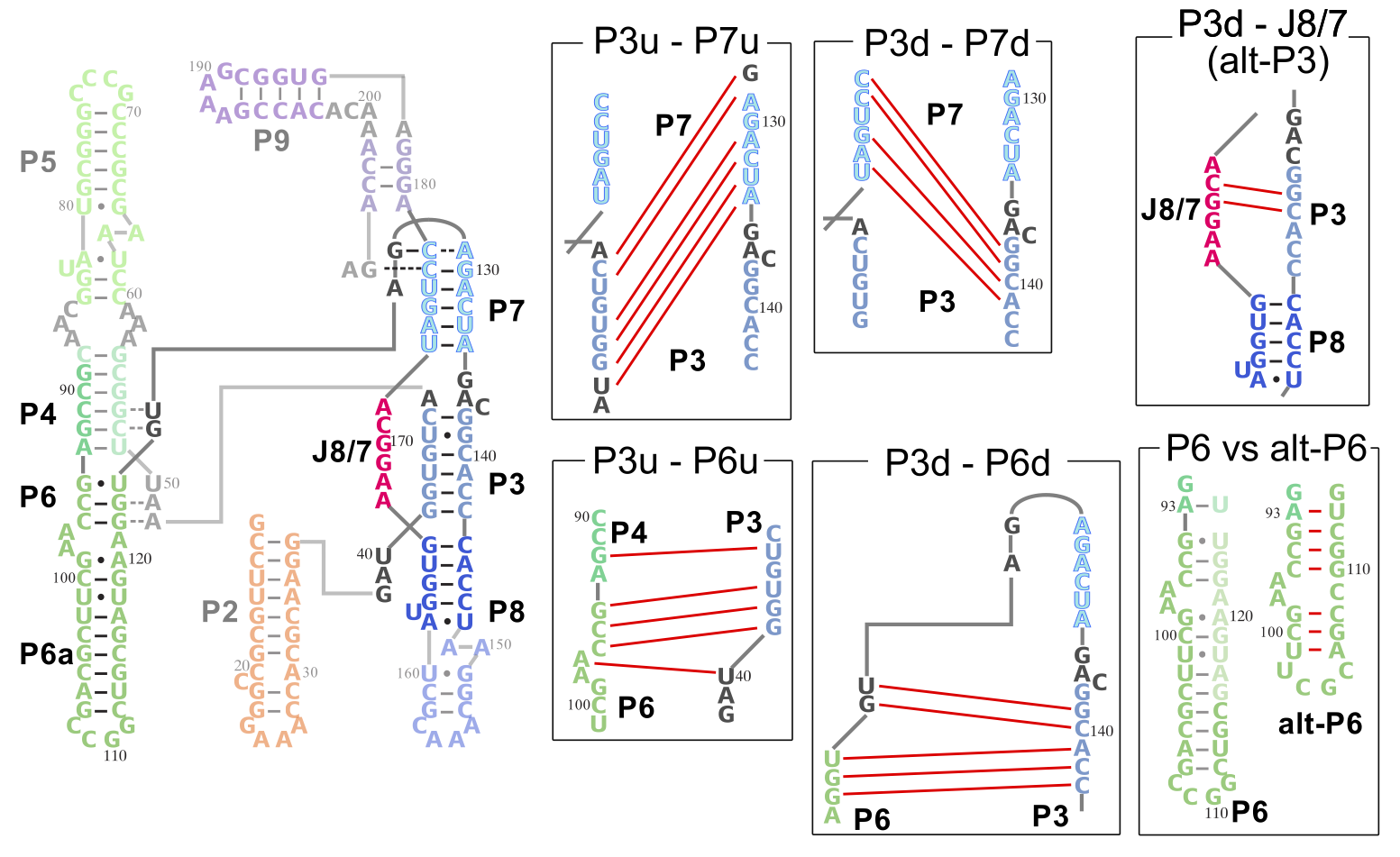}

\caption{
\label{fig:mispairing}
\textbf{Representative mispairing found in the simulations.}
The native secondary structure is shown on the left. Six frequently-observed mispairing patterns are shown in rectangles on the right. Each mispairing pattern is named with constituent strands, where each strand is distinguished as either up-stream (u) or down-stream (d) in helices.
}
\end{figure}

\begin{figure}
\includegraphics[width=0.5\textheight]{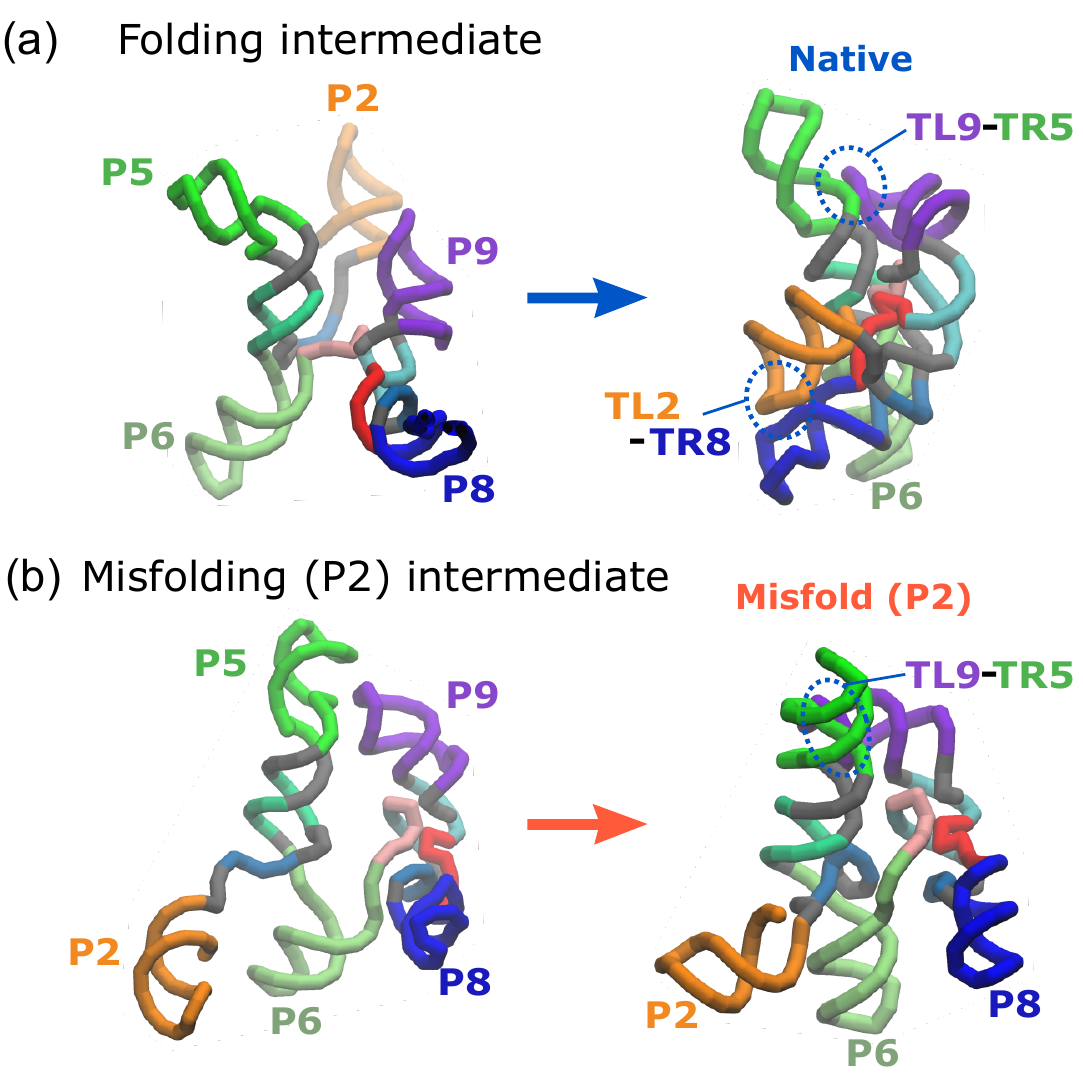}

\caption{
\label{fig:P2misfold_intermediate}
\textbf{Comparison between intermediate states leading to the P2 trapped state and correct folding.}
Two intermediate structures are extracted from trajectories (a) folding to the native structure and (b) misfolded (P2) state. The two intermediate states were visualized so that the relative positions of P5, P9 and P6 were approximately the same. In the correct folding (a, left), the P2 domain is positioned behind P5 and P9 helices, whereas in the misfolding (b, left) it comes to the front. Once the TL9-TR5 contact is formed in this situation, the P2 domain is unable to move backwards (b, right).
}

\end{figure}

\begin{figure}
\includegraphics[width=0.7\textheight]{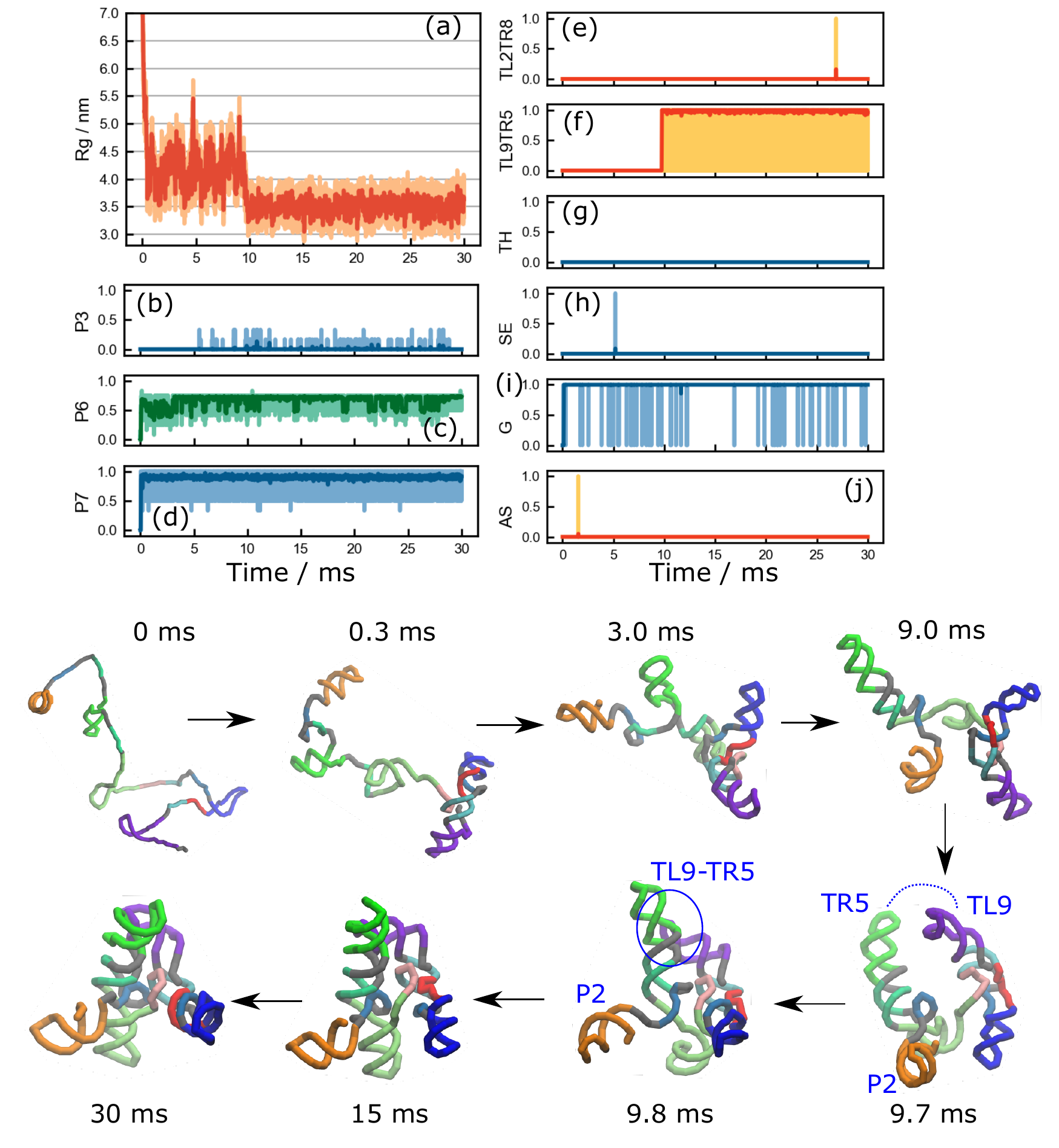}

\caption{
\label{fig:P2misfold_trapped}
\textbf{Time course of a trajectory kinetically trapped to Misfold (P2) state.}
One of the three trajectories in which the misfolded (P2) state was found. In this trajectory, at $t=$ 9.8 ms, the TL9-TR5 tertiary interaction formed when the P2 helix domain was mispositioned. The space between helices P3-P4-P5 (green) and P7-P8-P9 (blue, cyan, purple) is too narrow for the P2 helix to move back to the other side. Therefore the misfolded state remains until the end of the simulation.
}

\end{figure}

\begin{figure}
\includegraphics[width=0.7\textheight]{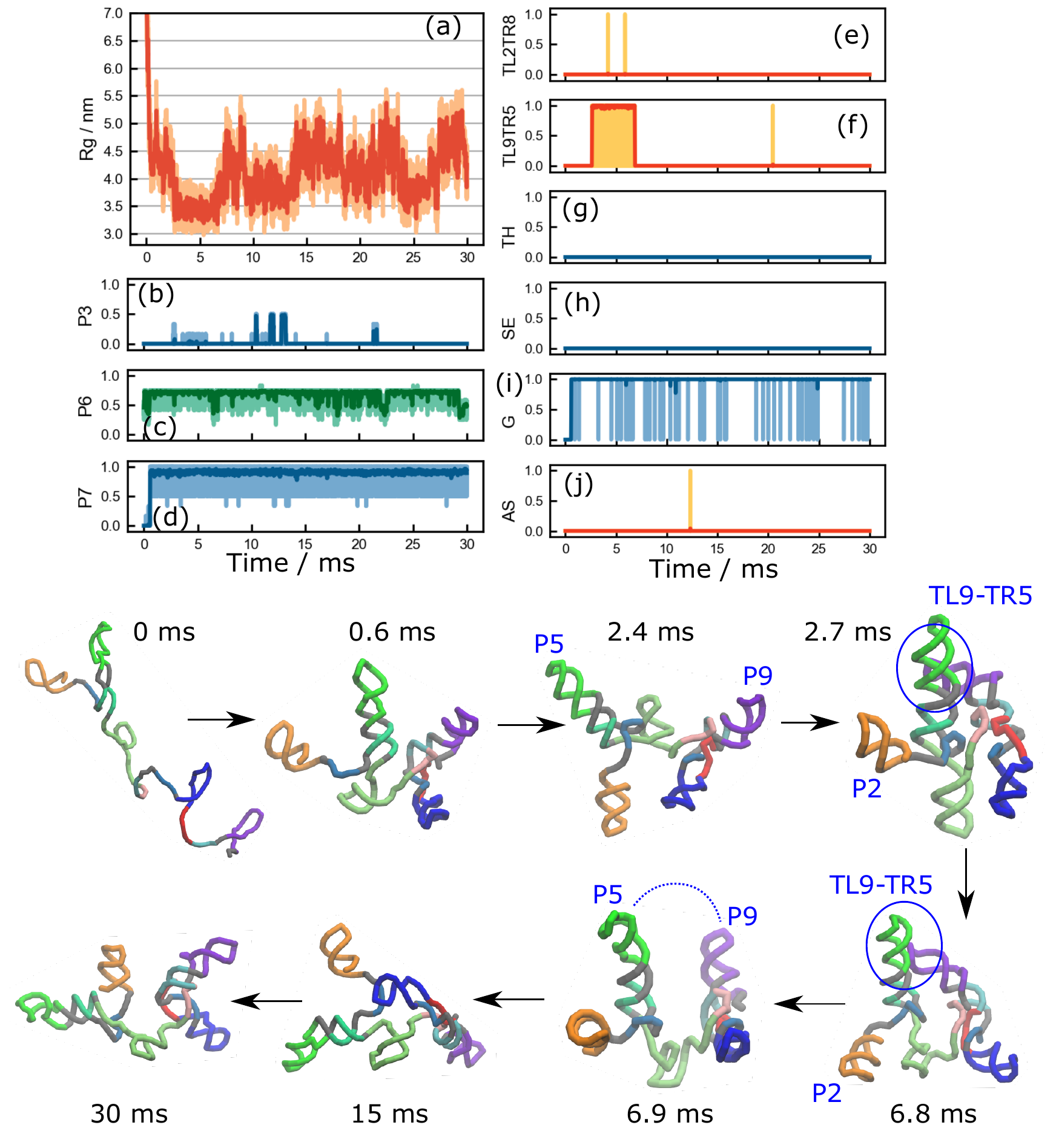}

\caption{
\label{fig:P2misfold_resolved}
\textbf{Escape from the Misfold (P2) state.}
This is the only trajectory where we observed escape from a topological trap.  Early in the trajectory ($t=$ 2.7 ms), the TL9-TR5 tertiary contact is formed while the P2 helix domain was in the misfolded position, in the same manner as in Figure \ref{fig:P2misfold_trapped}.  The ribozyme remained in the misfolded state until $t\sim$ 6.8 ms. Eventually, the TL9-TR5 interaction disengaged by the thermal fluctuation and the misfolded state was resolved. However, folding to the  native structure did not occur within the simulation time.
}
\end{figure}

\begin{figure}
\includegraphics[width=0.7\textheight]{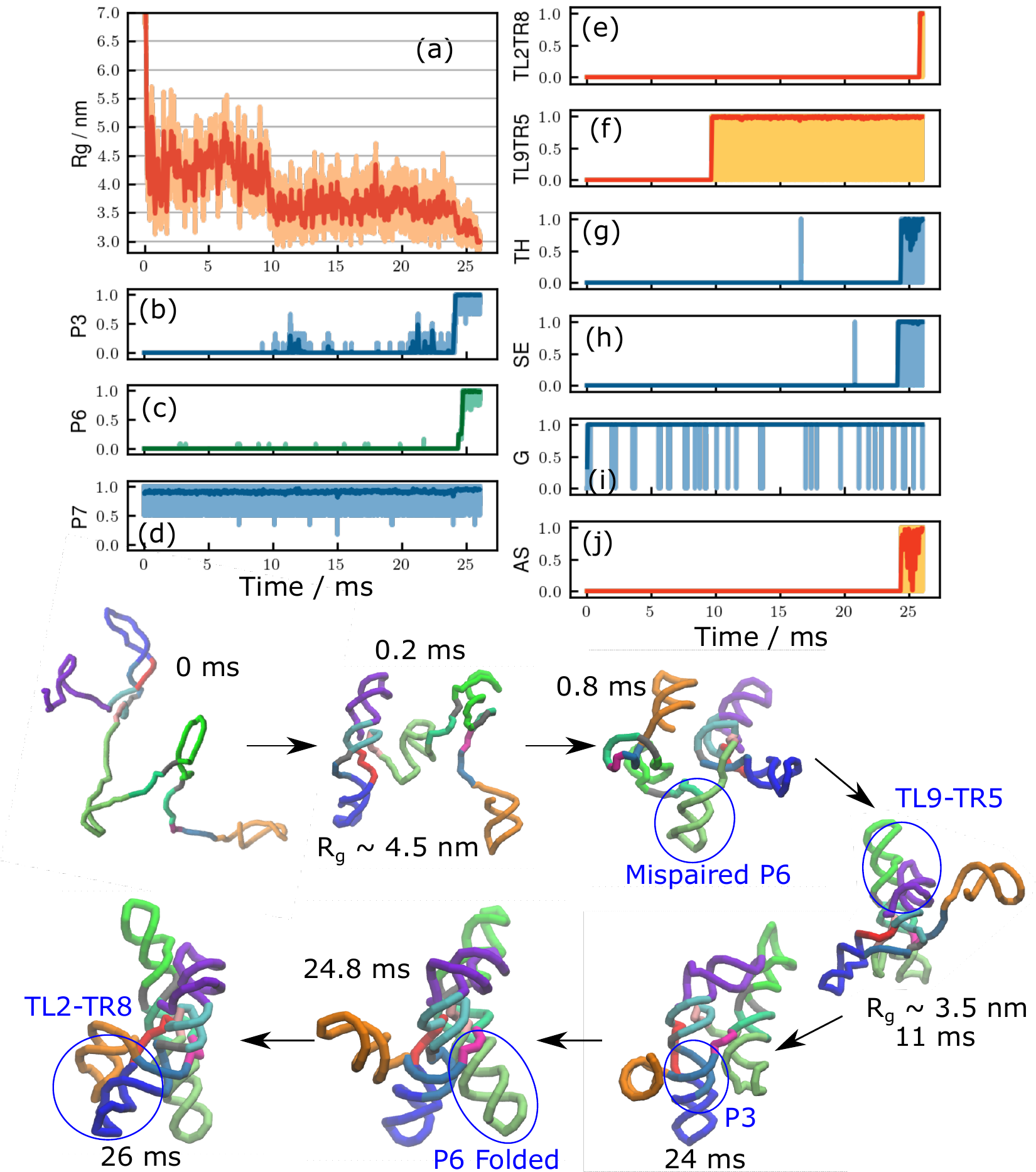}

\caption{
\label{fig:F033}
\textbf{Folding trajectory.} A trajectory that reached the folded state in a time scale larger than the one shown in the main text. A mispaired helix in P6 formed early ($t\sim0.8$ ms), which prevents it from folding to the native state rapidly. Eventually ($t\sim24$ ms), the misfolded helix was resolved, leading to the correct native fold. It should be pointed out that commitment to rapid specific collapse, defects in collapse but resolvable or trapping for arbitrarily long times in compact misfolded states occur roughly on the collapse time scale.
}
\end{figure}

\begin{figure}
\includegraphics[width=0.7\textheight]{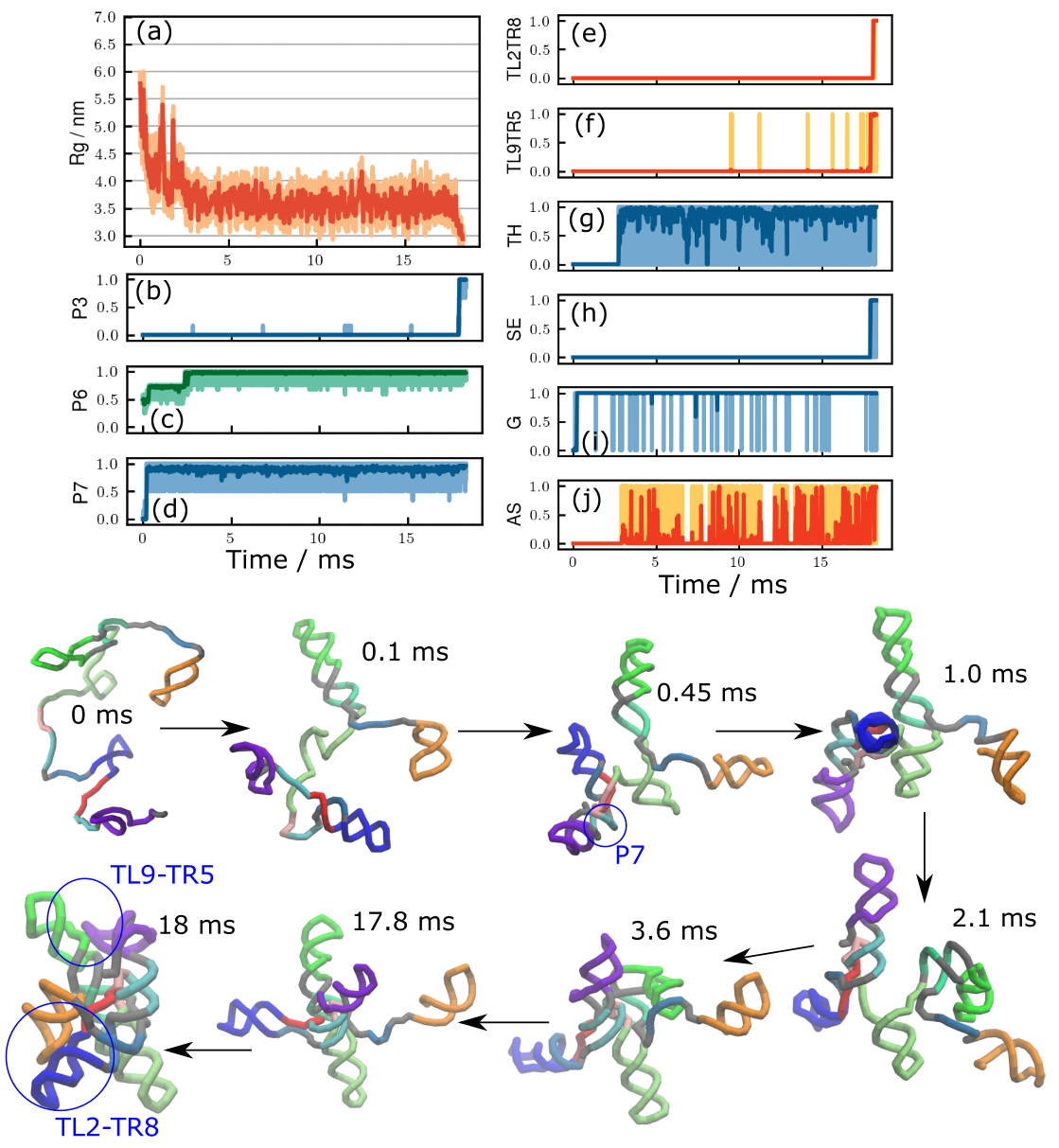}

\caption{
\label{fig:F057}
\textbf{Different routes to the folded state.} As with Figure~\ref{fig:F033}, this is another trajectory that reached the folded state in a time scale larger than the one shown in the main text. In this trajectory, helices P6 and P7 folded earlier ($t \sim 2$ ms), followed by the formation of the Triple Helix (TH). Two peripheral interactions, TL2-TR8 and TL9-TR5, needed longer times to search the counterparts. The folding was completed at $t\sim18$ ms. There is no typical folding trajectory.
}
\end{figure}

\begin{figure}
\includegraphics[width=0.7\textheight]{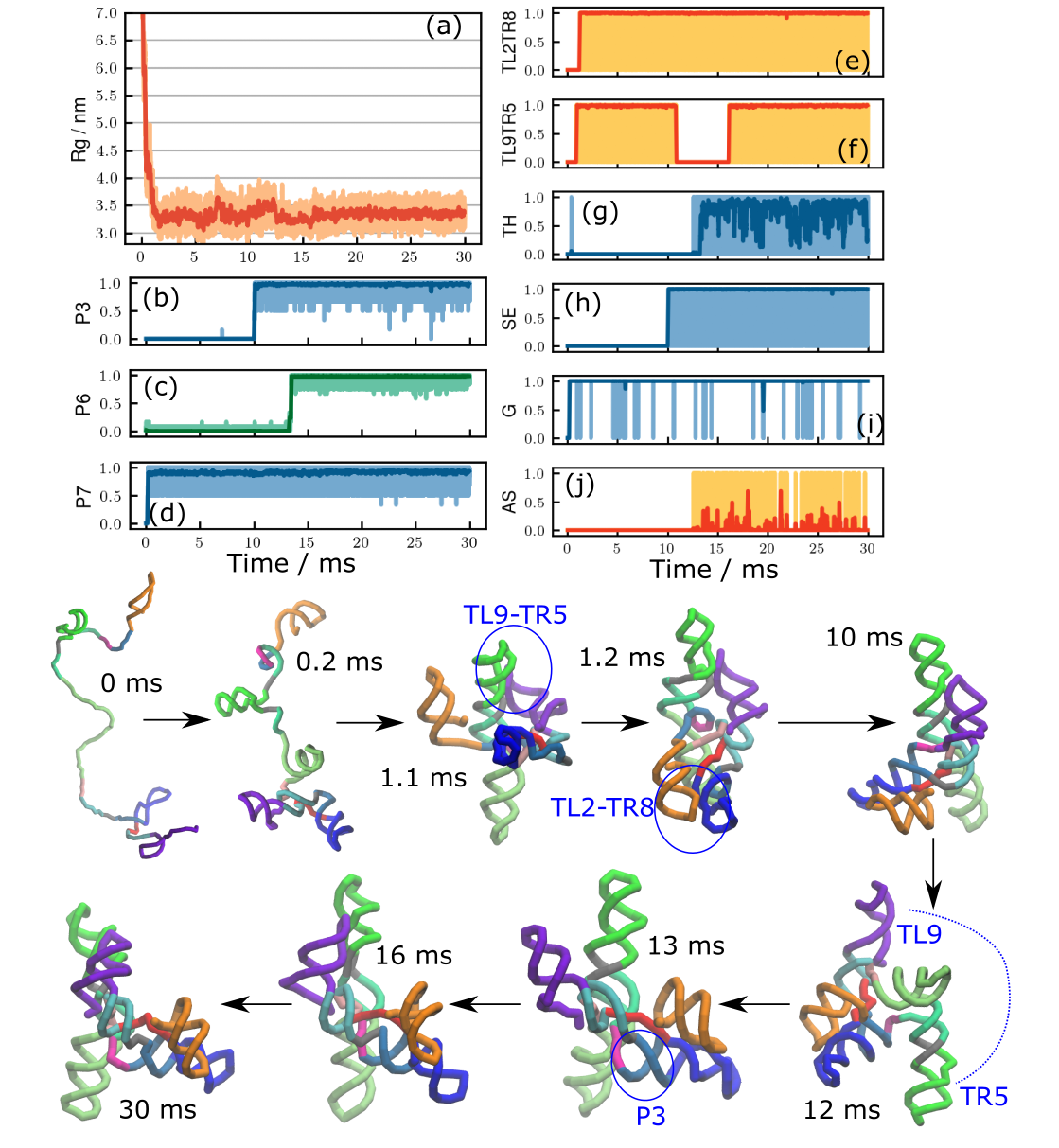}

\caption{
\label{fig:F042}
\textbf{Time course of a kinetically trapped trajectory.}
In this trajectory (also shown in Figure~\ref{fig:Fig4}), the mechanism of formation of a persistent misfolded state is revealed. Two key interactions in the peripheral regions, TL2-TR8 and TL9-TR5, formed at times $t < 1$ ms, result in the junction J8/7 with an incorrect topology (See Figure~\ref{fig:Fig7}b). Because the incorrect chain topology cannot be resolved unless both peripheral interactions unfold, the RNA stays in this misfolded state for the rest of the simulation time.  The trapped structure is an example of topological frustration.
}
\end{figure}

\begin{figure}
\includegraphics[width=0.7\textheight]{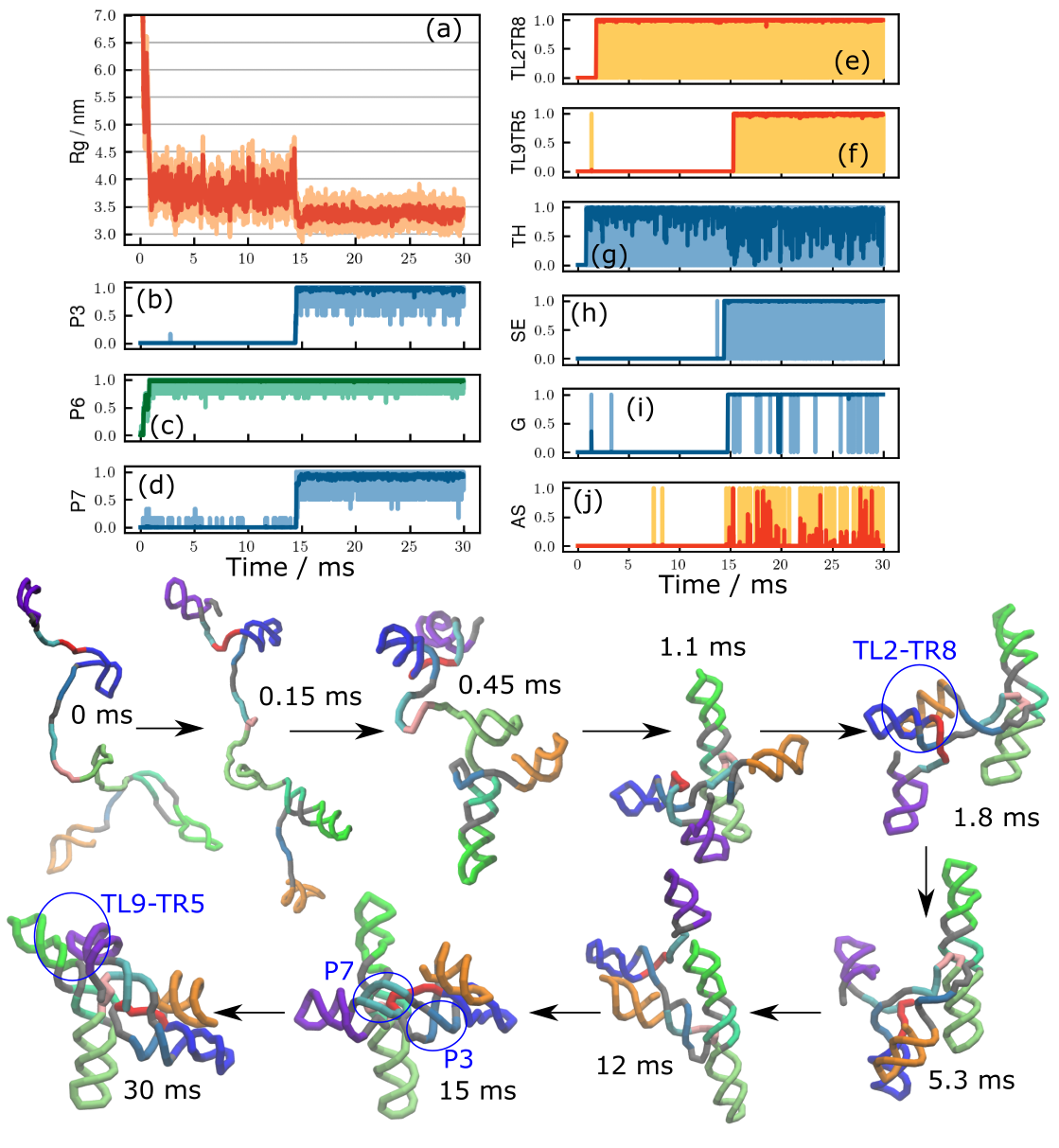}

\caption{
\label{fig:F090}
\textbf{Illustrating heterogeneity in the kinetically trapped trajectories.}
As with Figure~\ref{fig:F042}, this is another example of a kinetically trapped trajectory, that reached the same misfolded (J8/7) state. 
In this trajectory, it took a long time  ($t \sim 15$ ms) to form a compact intermediate state ($\Rg{} \sim 3.5$ nm) mainly because P3 and P7 helices did not form earlier. Also, unlike the previous example (Figure~\ref{fig:F042}), one of the peripheral interactions, TL9-TR5, formed in the later stage. 
}
\end{figure}

\begin{figure}
\includegraphics[width=0.6\textwidth]{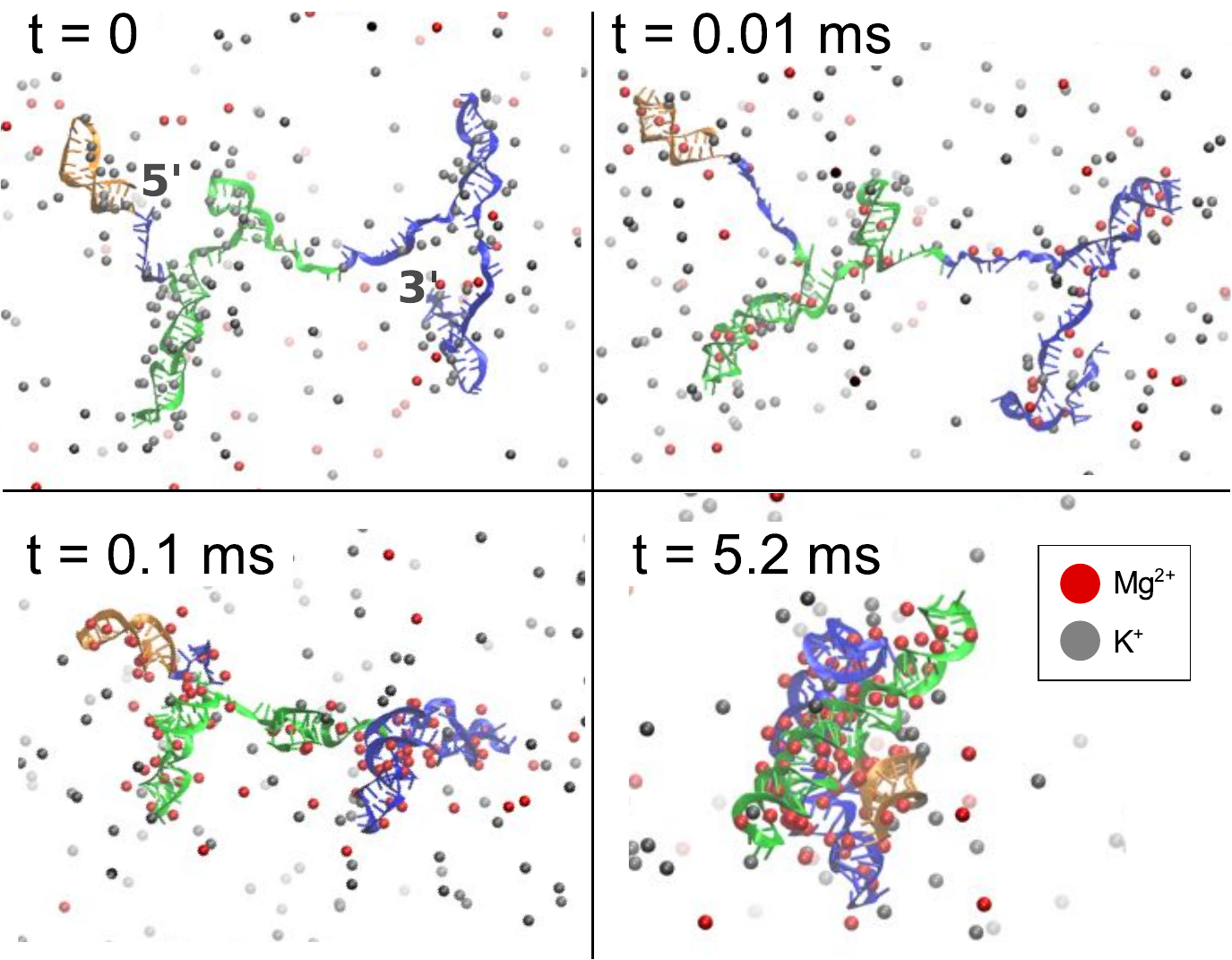}

\caption{
\label{fig:IonCondense_f002}
\textbf{Snapshots of ion condensations chosen from a folding trajectory.} The color scheme for ions is shown in the figure. For visual clarity, only cations in the vicinity of the RNA are shown. The size of spheres does not reflect actual volumes in the simulations. Although the majority of the ions near the folded RNA (last panel) are \Mg{}, there are detectable \K{} ions that also interact with the ribozyme in a site-specific manner. This implies that monovalent ions, even at low concentrations, could be present in the vicinity of the folded ribozyme. 
}
\end{figure}







\begin{figure}
\includegraphics[width=0.8\textwidth]{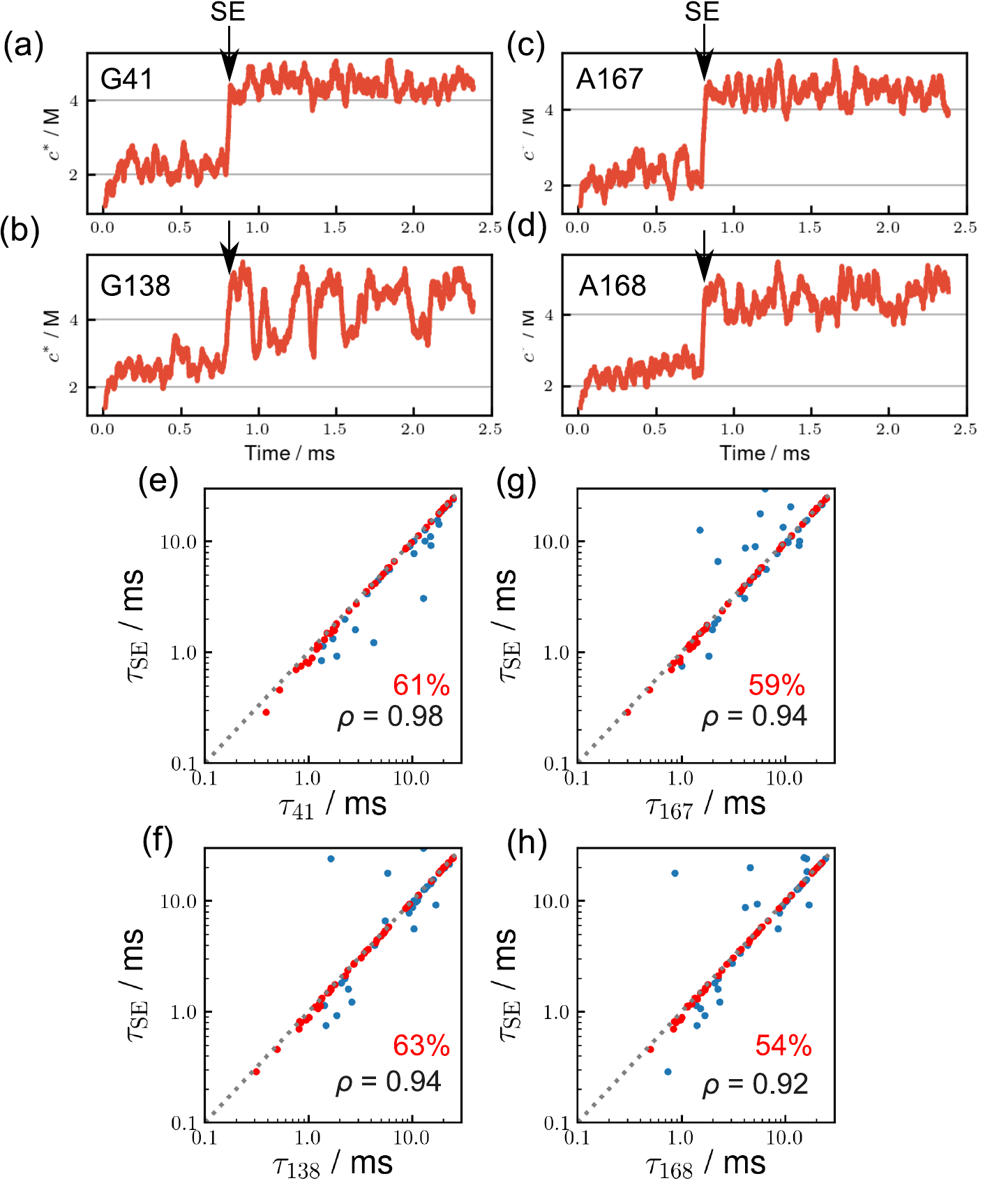}

\caption{
\label{fig:cor_SE}
\textbf{Time correlation between \Mg{} binding and the formation of Stack Exchange (SE).}  (a-d) Trajectories of \Mg{} binding, measured in terms of contact concentration, at (a) G41, (b) G138, (c) A167, and (d) A168 taken from the same simulation trajectory as Figure~\ref{fig:Fig3}. The folding time of SE is indicated by black arrows. (e-h) Scatter plots of the first passage times of \Mg{} binding to the same four nucleotides versus the first passage time of SE  formation. Cooperative binding of \Mg{} to these sites drives SE formation.
}
\end{figure}

\begin{figure}
\includegraphics[width=0.7\textwidth]{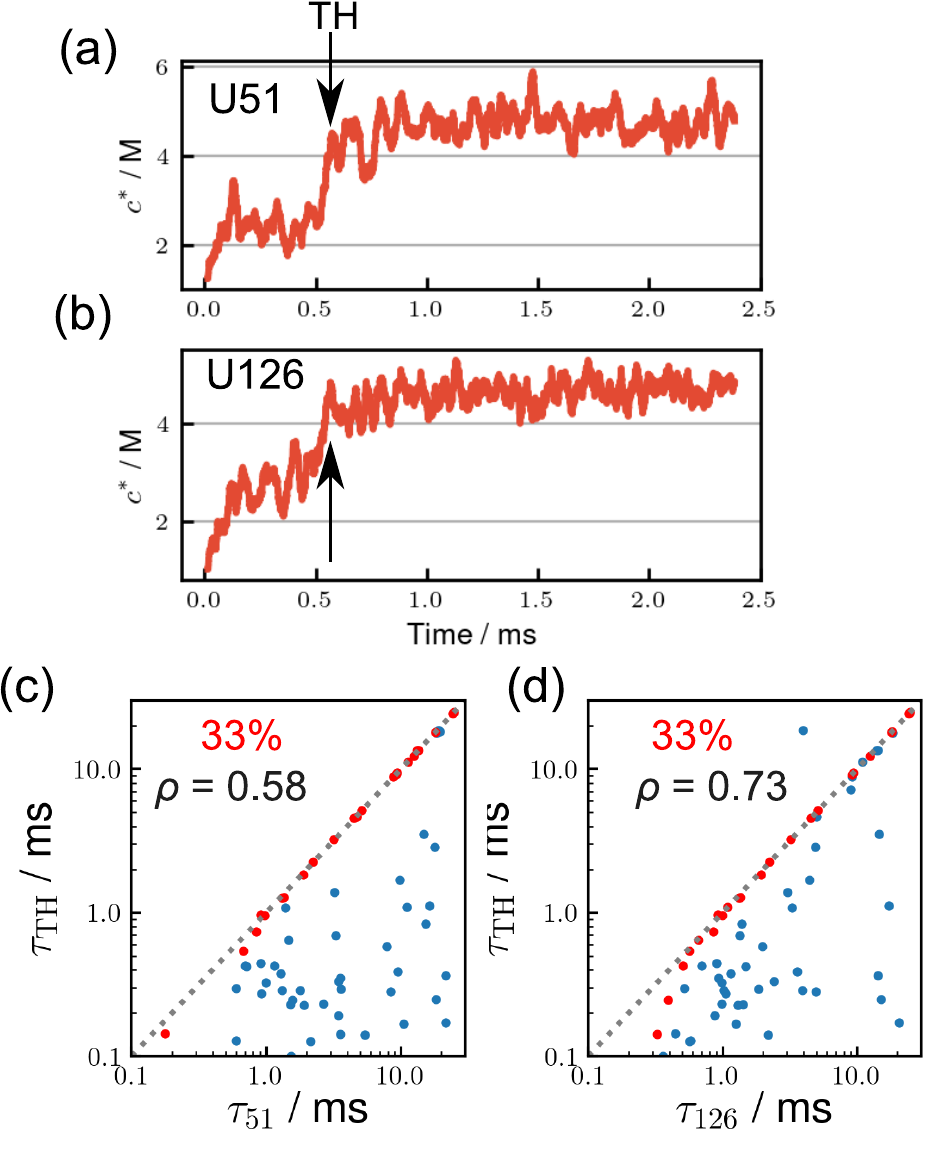}

\caption{
\label{fig:cor_TH}
\textbf{\Mg{} binding times to U51 and U126 and the formation of Triple Helix (TH).} 
\textbf{(a, b)} Trajectories of \Mg{} binding to (a) U51 and (b) U126 taken from the same simulation trajectory as Figure~\ref{fig:Fig3}. The folding time of TH is indicated by black arrows. 
\textbf{(c, d)} Scatter plots of the first passage times of \Mg{} binding to (c) U51 and (d) U126 versus the first passage time of the TH formation. Unlike the results in Figure~\ref{fig:cor_SE}, there are two modes of \Mg{} associations. In one class, TH formation and \Mg{} binding are concomitant. There are several trajectories in which \Mg{} association occurs after the TH formation (red dots). In this class, there is a broad distribution of \Mg{} binding times.
}
\end{figure}

\begin{figure}
\includegraphics[width=0.7\textwidth]{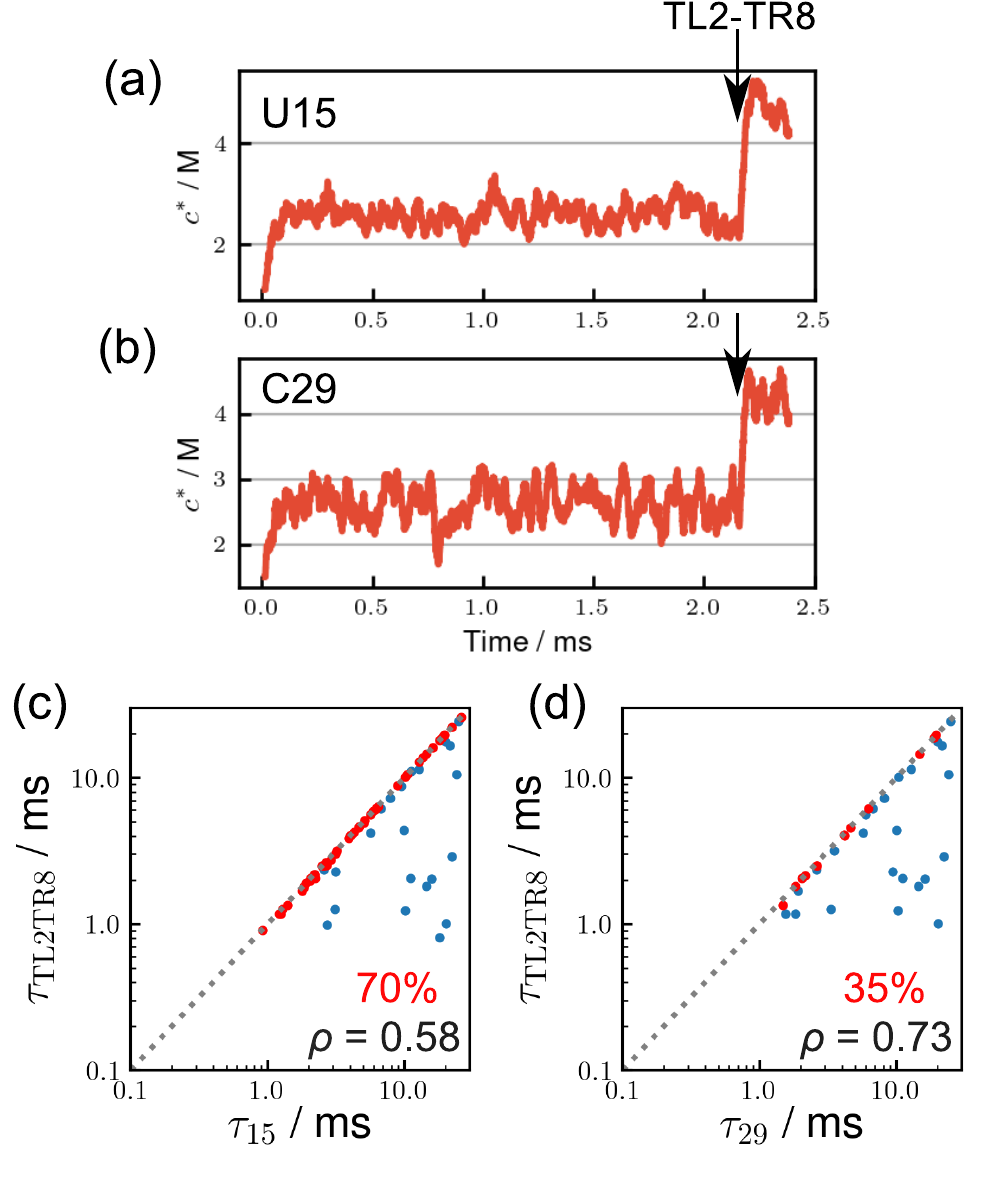}

\caption{
\label{fig:cor_TL2TR8}
\textbf{Same as Figure~\ref{fig:cor_TH} except association of \Mg{} to U15 and C29 and the formation of TL2-TR8 is shown.} 
\textbf{(a, b)} Trajectories of \Mg{} binding at (a) U15 and (b) C29 taken from the same simulation trajectory as Figure~\ref{fig:Fig3}. The folding time of TL2-TR8 is indicated by black arrows. 
\textbf{(c, d)} Scatter plots of the first passage times of \Mg{} binding to (c) U15 and (d) C29 versus the first passage time of the TL2-TR8 formation. Just as in Figure~\ref{fig:cor_SE}, we find correlated \Mg{} binding and TL2-TR8 formation, as well as structure formation followed by \Mg{} binding.
}
\end{figure}

\begin{figure}
\includegraphics[width=0.8\textwidth]{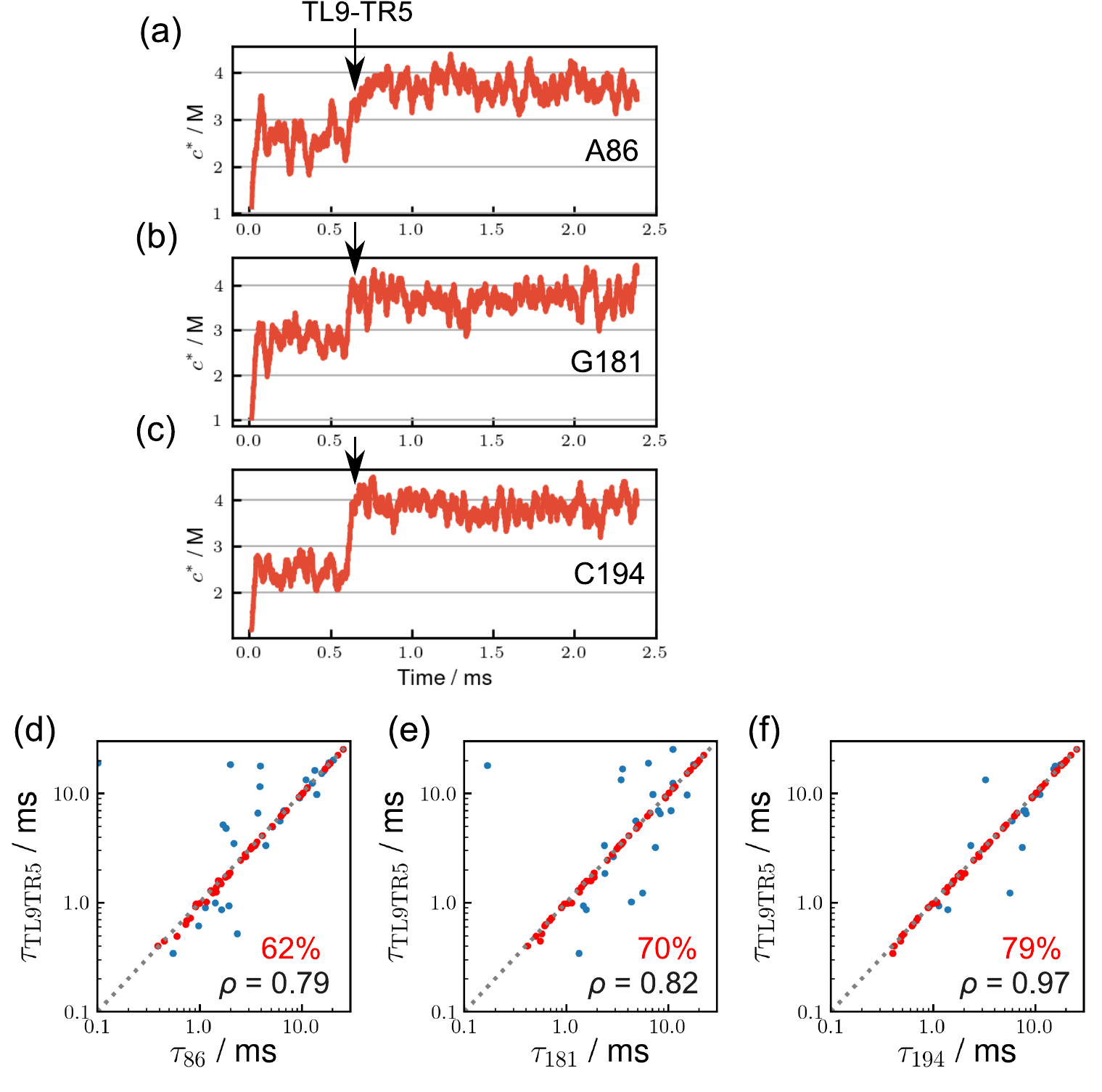}

\caption{
\label{fig:cor_TL9TR5}
\textbf{Coordinated formation of TL9-TR5 and \Mg{} binding to A86, G181, and C194.}
\textbf{(a-c)} \Mg{} binding to (a) A86, (b) G181, and (c) C194 taken from the same simulation trajectory as Figure~\ref{fig:Fig3}. The black arrows show the folding time of TL9-TR5.
\textbf{(d-f)} Scatter plots of the first passage times of \Mg{} binding to (d) A86, (e) G181, and (f) C194 versus the first passage time of the TL9-TR5 formation. In the majority of these trajectories, there is a high degree of specific coordinated \Mg{} binding and TL9-TR5 formation. 
}
\end{figure}

\begin{figure}
\includegraphics[width=0.8\textwidth]{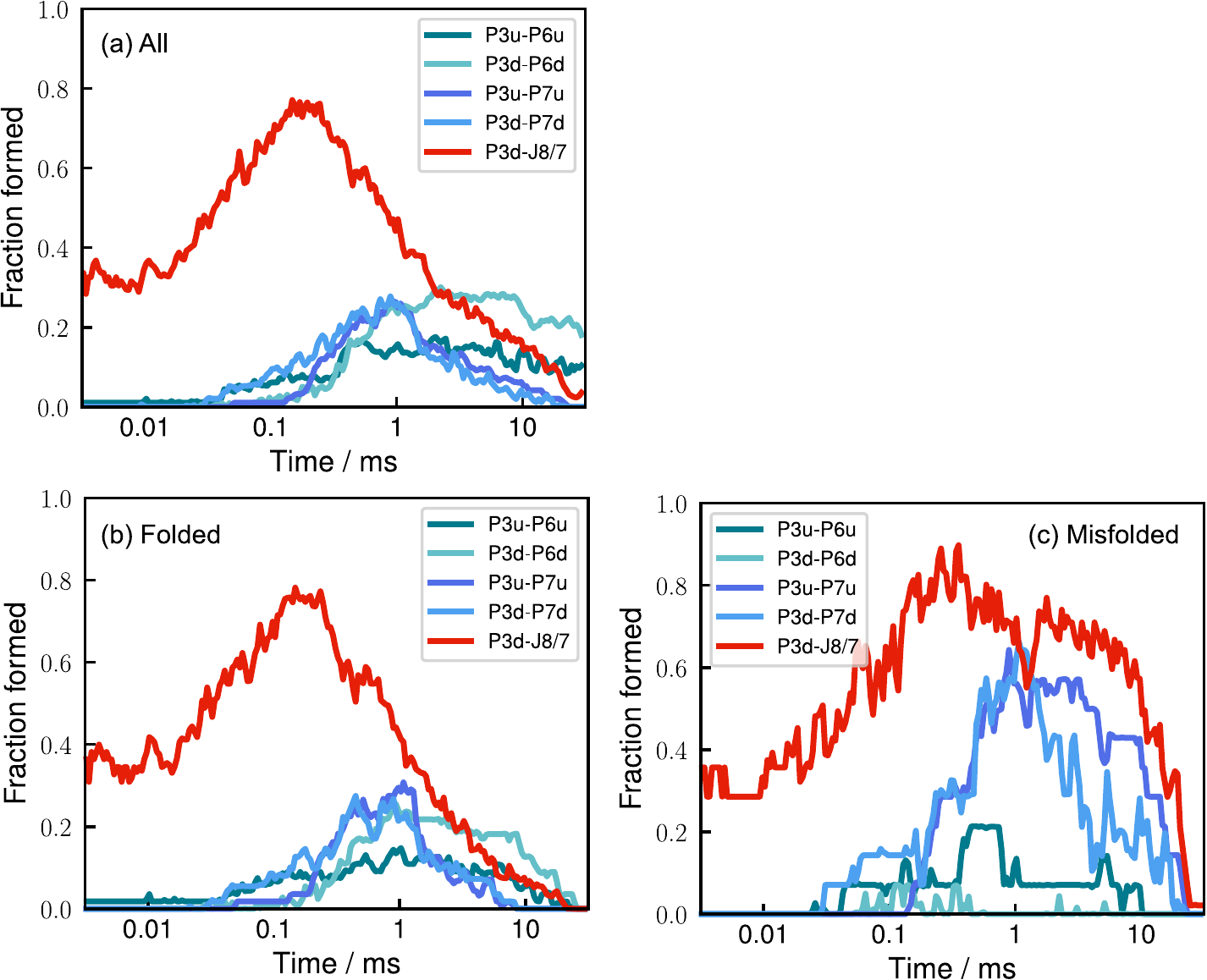}

\caption{
\label{fig:nnpair_fraction}
\textbf{Formation of the five mispaired helices, which slows the folding of the ribozyme.}
\textbf{(a)} Time-dependent fractions of mispaired helices averaged over all the trajectories. 
\textbf{(b, c)} Same as (a) except those correspond to trajectories that reach the folded state (b), and the misfolded (J8/7) state (c). See Figure~\ref{fig:mispairing} for the notation of the mispaired helices.
}
\end{figure}